\documentclass[journal]{IEEEtran}
\usepackage{cite} 
\usepackage{url}  
\usepackage{ifthen}  
\usepackage[utf8]{inputenc} 
\usepackage{amsfonts}
\usepackage{amssymb}

\usepackage{amsthm}
\usepackage{amsmath}
\usepackage{times}
\usepackage{latexsym}
\usepackage{bm}
\usepackage{amssymb}
\usepackage{stfloats}
\usepackage{cases}
\usepackage{array}
\usepackage{setspace}
\usepackage{fancyhdr}
\usepackage{multirow}
\usepackage{algorithmic}

\usepackage{psfrag}
\usepackage{subfigure}
\usepackage{url}
\usepackage{epsfig}
\usepackage{caption2}
\usepackage{wrapfig}
\usepackage{makecell}
\usepackage{color}
\usepackage{algorithm}
\usepackage{algorithmic}
\fnbelowfloat
\ifCLASSOPTIONcaptionsoff
 \usepackage[nomarkers]{endfloat}
 let\MYoriglatexcaption\caption
 \renewcommand{\caption}[2][\relax]{\MYoriglatexcaption[#2]{#2}}
\fi

\begin{document}

\title{Spectrum Sensing under Spectrum Misuse Behaviors: A Multi-Hypothesis Test Perspective}

\author{Linyuan Zhang, Guoru Ding, \IEEEmembership{Senior Member, IEEE},\\ Qihui Wu, \IEEEmembership{Senior Member, IEEE}, and Zhu Han,~\IEEEmembership{Fellow,~IEEE}

\thanks{This work is supported by the National Natural Science Foundation of China (Grant No. 61501510 and No. 61631020), Natural Science Foundation of Jiangsu Province (Grant No. BK20150717), Natural Science Foundation for Distinguished Young Scholars of Jiangsu Province under Grant (No. BK20160034), and in part by US NSF CNS-1717454, CNS-1731424, CNS-1702850, CNS-1646607, ECCS-1547201, CMMI-1434789, CNS-1443917, and ECCS-1405121.}

\thanks{L. Zhang and G. Ding are with the College of Communications Engineering, Army Engineering University of PLA, Nanjing 210007, China (e-mail: zhanglinyuan5@163.com, dr.guoru.ding@ieee.org). G. Ding is also with the National Mobile Communications Research Laboratory, Southeast University, Nanjing 210018, China.}

\thanks{Q. Wu is with the Department of Electronics and Information Engineering, Nanjing University of Aeronautics and Astronautics, Nanjing 210007, China (email: wuqihui2014@sina.com).}

\thanks{Z. Han is with the University of Houston, Houston, TX 77004 USA (e-mail:zhan2@uh.edu), and also with the Department of Computer Science and Engineering, Kyung Hee University, Seoul, South Korea.}}

\maketitle
\newtheorem*{rem}{Remark}
\newtheorem{lem}{Lemma}
\newtheorem{pro}{Proposition}
\newtheorem{thm}{Theorem}

\begin{abstract}
Spectrum misuse behaviors, brought either by illegitimate access or by rogue power emission, endanger the legitimate communication and deteriorate the spectrum usage environment. In this paper, our aim is to detect whether the spectrum band is occupied, and if it is occupied, recognize whether the misuse behavior exists. One vital challenge is that the legitimate spectrum exploitation and misuse behaviors probabilistically coexist and the illegitimate user (IU) may act in an intermittent and fast-changing manner, which brings about much uncertainty for spectrum sensing. To tackle it, we firstly formulate the spectrum sensing problems under illegitimate access and rogue power emission as a uniform ternary hypothesis test. Then, we develop a novel test criterion, named the generalized multi-hypothesis Neyman-Pearson (GMNP) criterion. Following the criterion, we derive two test rules based on the generalized likelihood ratio test (GLRT) and the Rao test, respectively, whose asymptotic performances are analyzed and an upper bound is also given. Furthermore, a cooperative spectrum sensing scheme is designed based on the global GMNP criterion to further improve the detection performances. In addition, extensive simulations are provided to verify the proposed schemes' performance under various parameter configurations.
\end{abstract}


\section{Introduction}
\subsection{Background and Motivation}
\begin{figure}[t]
\centering
\subfigure[Illegitimate access]{
\includegraphics[width=0.45\linewidth]{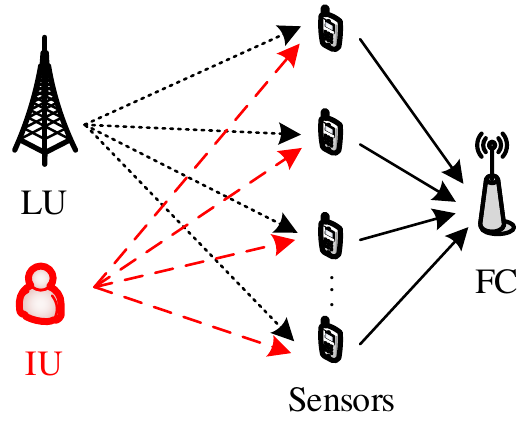}}
\subfigure[Rogue power emission]{
\includegraphics[width=0.45\linewidth]{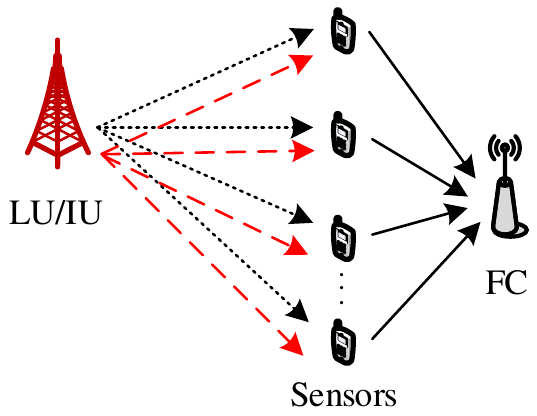}}
\caption{Spectrum misuse behaviors and a distributed detection framework. Dotted lines denote the received signal of sensors due to the probabilistic emission behaviors of the IU and the legitimate user (LU), while the solid lines denote that sensors report decisions to the fusion center (FC).}
\label{Fig-system}
\end{figure}
Due to the explosive growth of wireless devices and services, the scarceness of spectrum resources has become more and more serious. Dynamic spectrum sharing has been well recognized as a promising approach to improve the spectrum utilization and relieve the spectrum shortage \cite{Mitola,Haykin,Dynamic_spectrum}. However, it may put the future at risk brought by various attacks \cite{CST,Ding_Tcom}, among which spectrum misuse behaviors have received increasingly wide attention. Generally, spectrum misuse behaviors refer to the manner of exploiting spectrum that violates the spectrum regulations. As illustrated in Fig. \ref{Fig-system}, two typical spectrum misuse behaviors are as follows:
\begin{itemize}
  \item \emph{Illegitimate access} launched by external attackers, just as shown in Fig. \ref{Fig-system} (a). One example is the primary user emulation (PUE) attack \cite{Chen_2008,GLobecom_2012,Dogfight_1,Dogfight_2} that launches the denial of service (DoS) attack, where the illegitimate user emulates the characteristics of the primary user's signal when the sensors conduct spectrum sensing in order to prevent the secondary users from accessing the idle spectrum \cite{nonpa_Bayesian,Belief_PUE,ALDO,PUEA_model}. Another example is the fake or rogue access point (AP) that disguises as an authorized AP by spoofing the authorized AP's medium access control (MAC) address in order to lure users to connect to them \cite{Clock_Skews}\cite{Permits_mobihoc}. In addition, the illegitimate access can be done indirectly through launching data falsification attacks in the cooperative spectrum sensing \cite{Lin_access}\cite{Nie_access}, where the global decisions are misled and as a result, collisions between secondary users and the primary user increase and attackers illegitimately access the idle channel \cite{Attack_preve,Thwart}.
  \item\emph{Rogue power emission} taken by internal attackers, just as shown in Fig. \ref{Fig-system} (b). In some cases, a selfish attacker may break the upper bound of the transmit power constraint for a higher data rate \cite{See_something}\cite{SpecWathch}. Similarly, some malicious attackers may intentionally apply for spectrum exploitation with a large coverage, quoting a high transmit power, but in fact they work with a low transmit power, which decreases the spectrum efficiency of spatial reuse \cite{rogue_mass2012}. The internal attackers have obtained the authorization of the usage of the spectrum, and as a result, this kind of misuse behaviors is stealthier.
\end{itemize}

These spectrum misuse behaviors lead to serious interference with the legitimate communication and disordered spectrum usage environment, which poses a fundamental requirement on the detection of these misuse behaviors.
\subsection{Related Work}
Spectrum misuse behaviors pose direct influences on the channel states, while spectrum sensing plays a vital role in obtaining the real-time channel states. Further, considering the high feasibility, spectrum sensing based methods have been exploited in the detection of spectrum misuse behaviors \cite{ALDO,Yao_css,Three_fingerpints,See_something,Ad_hoc}.
Specifically, the detection problem of the illegitimate access is formulated as a statistical significance test between the normal usage and the abnormal one in \cite{Ad_hoc}, where the normal usage is defined as being no more than one transmitter working in each portion (e.g., channel) of the spectrum. But, it can't find out the case that only the IU works. To tackle this problem, \cite{ALDO} introduces the spectrum policy maker to obtain the factual state of the LU, based on which the detection is divided into two cases, i.e., when the LU isn't working and when the LU is working. Especially, when the existence of the IU taking illegitimate access is prior known, an optimal combining scheme is proposed for cooperative spectrum sensing (CSS) in \cite{Yao_css}. Differently, considering the assumption that no more than one user, either the LU or the IU, works on a spectrum band of interest during one slot, \cite{Three_fingerpints} proposes a fingerprints-power-belief based noncentral detection algorithm through making power estimation and calculating the compatibility in a distributed manner. On the other side, few works have been done for the rogue power emission problem.
In \cite{See_something}, crowdsourced enforcement, where a crowd of mobile users collaterally make detection, is exploited to find out the spectrum misuse behavior and the related characteristics, e.g., the signal strength, are analyzed.

In fact, one basic but vital problem is that the detectors generally have no ground truth of either the legitimate exploitation or misuse behaviors. Hence, in this paper, we focus on two new features about the sensing model:
\begin{itemize}
  \item \emph{From binary-hypothesis to multi-hypothesis}: We not only identify whether the spectrum band is occupied or not, but also have to make clear whether the band, if occupied, is occupied legitimately or illegitimately. Due to lack of the ground truth about the LU's behaviors, the traditionally binary-hypothesis test model is inaccurate to characterize the new problem any more. Hence, considering the probabilistic coexistence of the IU and the LU, it is essential to formulate the detection problem as a multi-hypothesis test problem.
  \item \emph{From simple hypothesis to composite hypothesis}: From the perspective of protecting itself from being detected, the IU may behave in an intermittent and fast-changing manner. As a result, it is hard to obtain the accurate parameters of the IU, such as the transmit power and channel gains, which means that the composite hypotheses with a family of distributions are needed to formulate the behaviors instead of simple ones with a single distribution.
\end{itemize}

To the best of the authors' knowledge, there is no work considering the detection of spectrum misuse behaviors as a multi-hypothesis test consisting of simple hypotheses and composite hypotheses, though multiple hypotheses and composite hypotheses have been separately considered in current work about spectrum sensing.
In \cite{Gaofeifei}, the case that the licensed user works on more than one discrete transmit power levels is considered and a multi-hypothesis test is made to detect the existence of the signal and recognize the power levels based on the maximum a posterior rule. Further, for the same multi-power-level case, a sequential detection is studied based on the modified Neyman-Pearson (NP) rule in \cite{Sequ_multi}. As the transmit power levels are deterministic and known for the sensors, all hypotheses are simple. However, when some key parameters are unknown, the hypothesis becomes composite. In \cite{Genera_Rao}, a two-sided parameter testing problem is analyzed, where the user's transmitting power and the position are unknown, and a generalized Rao test is derived for the binary hypothesis test consisting of a simple hypothesis and a composite one.
Nevertheless, this paper formulates a different multiple hypothesis test, which consists of composite hypotheses and simple ones, and the problem is complex and hard to obtain an optimal scheme.


\subsection{Contributions}
In this paper, a distributed detection framework is built, where every sensor makes spectrum sensing and local decision, and based on the decisions received from the sensors, the FC makes data fusion and global decision, just as shown in Fig. \ref{Fig-system}. Our aim is to detect whether the channel is occupied, and if it is occupied, recognize whether the IU exists. The main contributions are summarized as follows:
\begin{itemize}
  \item We firstly formulate the spectrum sensing problem under illegitimate access and rogue power emission as multi-hypothesis tests, respectively, where the states without the IU are modeled as simple hypotheses and the states with the IU are modeled as composite ones due to the unknown characteristics about the IU, and  give a uniform model for the two cases, i.e., a ternary hypothesis test, through certain transformation to facilitate the analysis and processing.
  \item We propose a generalized multi-hypothesis Neyman-Pearson (GMNP) criterion and derive the test rule. Specifically, in the GMNP criterion, the detection probability about spectrum misuse behaviors is maximized under two constraints about the detection probabilities of states without the IU. Following the criterion, two test rules are derived based on the GLRT and the Rao test, respectively. In particular, due to the multi-hypothesis characteristic, the overlapping problem of decision regions is analyzed and the solution is given. To evaluate the proposed schemes, the asymptotic performance is derived and an upper bound of the detection performance is also given.
  \item We design a cooperative spectrum sensing (CSS) scheme for the ternary hypothesis test based on the global GMNP test criterion. Due to the unknown detection performance when the IU exists, the criterion is further transformed, based on which a data fusion-based CSS algorithm is developed.
  \item We present in-depth simulations to verify the detection performance versus the unknown parameter, the number of sensors, and the number of samples in two scenarios: sensing with one single sensor and sensing with multiple sensors.
\end{itemize}

Notably, the proposed method can handle with both illegitimate access and rogue power control from the perspective of spectrum sensing. Considering that current works focus on either the illegitimate access or rogue power control, our work shows high efficiency.

\subsection{Organization and Notation}
The rest of this paper is organized as follows. The system model under spectrum misuse behaviors is formulated in Section II. Section III formulates the GMNP criterion, derives the test rule, and gives an upper bound of the detection performance. Cooperative spectrum sensing is analyzed in Section IV. Simulation is given in Section V, followed by the conclusion in Section VI. The proofs of all theorems are given in the appendix. To facilitate the reading, the key notations are summarized in Table I.
\begin{table}[!t]
\begin{center}
\caption{Key notations and symbols used in this paper.}
\begin{tabular}{|c|l|}
\hline
Symbol&Definition
\\\hline
$LU$&The legitimate user
\\\hline
${IU}$&The illegitimate user
\\\hline
$\Pr(\cdot )$&the probability of certain event
\\\hline
$F(\cdot)$&\makecell[l]{the probability distribution function of\\ certain parameter}
\\\hline
$p(\cdot)$&\makecell[l]{the probability density function of certain\\ parameter}
\\\hline
${\mathcal{H}_i}$ & the hypothesis $\mathcal{H}_i,~i\in\{0,1,2\}$
\\ \hline
${\mathcal{R}_i}$ & the corresponding decision region of $\mathcal{H}_i$
\\ \hline
$N$ & the number of samples
\\ \hline
$K$ & the number of sensors
\\ \hline
\end{tabular}
\end{center}
\label{tab:notion}
\end{table}

\section{System Model Under Spectrum Misuse Behaviors}

\subsection{Problem Formulation}
Conventionally, to detect whether a LU is transmitting signals in a given spectrum band, a binary hypotheses test is carried out between the null hypothesis $\mathcal{L}_0$ (the LU is absent) and the alternative hypothesis $\mathcal{L}_1$ (the LU is present) \cite{spectrum_sensing}:
\begin{equation}
\begin{gathered}
  {\mathcal{L}_0}:{y_L}(t) = n(t), \hfill \\
  {\mathcal{L}_1}:{y_L}(t) = \sqrt {{P_s}}s(t) + n(t), \hfill \\
\end{gathered}
\label{Eq_rogue0}
\end{equation}
where ${y_L}(t)$ is the received signal of a sensor in time slot $t$, $s(t)$ is the signal transmitted by the LU, $P_s$ is the received power that is related with the transmit power and the path loss and invariable during the observation time, and $n(t)$ is the additive Gaussian white noise with the variance $\sigma_n^2$.

Further, to facilitate the analysis, in the following, $s(t)$ is assumed to be an independent and identically distributed (i.i.d.) random process with zero mean and unit variance, i.e., $s(t) \sim \mathcal{N}(0, 1)$ \cite{Yao_css}\cite{sens_throu_trad}. Here, $P_s$ and $\sigma_n^2$ are assumed to be prior known, and it is considered from two aspects: it is wise for the LU to provide some information (such as the transmit power and statistical channel state) to assist sensors of well finding out the mis-behaved IU \cite{ALDO}; considering the LU's regularity compared with the IU's burstiness, it is feasible to estimate the received power $P_s$ of the LU and the noise $\sigma_n^2$ over a long observation. Hence, in ${\mathcal{L}_1}$, ${y_L}(t) \sim \mathcal{N}(0, P_s+\sigma_n^2)$.
Over a sensing slot $T$, the sensor obtains $N$ samples, $\mathbf{y}=(y_0,y_1,...,y_N)$, based on which the test is done.

Next, we analyze two cases of spectrum misuse behaviors, illegitimate access and rogue power control, respectively.

\subsubsection{Illegitimate access}
In this case, the IU is an unauthorized user taking illegitimate access, and another binary hypotheses test about the IU can be modeled as follows:
\begin{equation}
\begin{gathered}
  {\mathcal{I}_0}:{y_I(t)} = 0, \text{if~no~IU~exists;} \hfill \\
  {\mathcal{I}_1}:{y_I(t)} = \sqrt {{P_x}} x(t), \text{if~an~IU~exists.} \hfill \\
\end{gathered}
\end{equation}
where $y_I(t)$ is the illegitimate component in the received signal, and $x(t)$ is the signal transmitted by the IU with the received power $P_x$. As the IU always imitates the signal characteristics of the LU to hide itself from being detected, $x(t)$ is also assumed to be an i.i.d. random process with zero mean and unit variance, i.e., $x(t) \sim \mathcal{N}(0, 1)$. In addition, as the IU generally works abruptly in a non-cooperative manner with the detection system in order to better hide itself from being detected, $P_x$ is hard to obtain. Nevertheless, considering the short sensing time and the ignorable, $P_x$ is assumed to be invariable during one sensing slot.
Furthermore, let $P(\mathcal{L}_i)$ and $P(\mathcal{I}_j)$, $i,j \in\{0,1\}$, denote the prior probabilities about the existence of the LU and the IU, respectively. We have
\begin{equation}
\Pr ({\mathcal{L}_i},{\mathcal{I}_i}) = P({\mathcal{L}_i})\Pr ({\mathcal{I}_i}|{\mathcal{L}_i}).
\label{Eq_coexistence}
\end{equation}

\begin{rem}
The dependence of the IU's states and the LU's states varies with different goals and the prior information.
In the case that the IU aims to well exploit the spectrum for free, e.g., the PUE attack, it prefers to work when the LU is absent and $\Pr ({\mathcal{I}_1}|{\mathcal{L}_0})$ may be higher than $\Pr({\mathcal{I}_1}|\mathcal{L}_1)$. While in another case that the IU's goal is to deteriorate normal communication of the LU, e.g., the jamming attack, $\Pr ({\mathcal{I}_1}|{\mathcal{L}_1})$ may be higher than $\Pr({\mathcal{I}_1}|\mathcal{L}_0)$. However, it is generally hard to obtain accurate information about the states of the LU for the IU. In an extreme case that there is no information about the real state of the LU, the IU has to work completely randomly and independently, i.e., $\Pr ({\mathcal{L}_i},{\mathcal{I}_i})=P({\mathcal{I}_i})P({\mathcal{L}_i})$.
\end{rem}
\subsubsection{Rogue power control}
In this case, the IU is the authorized user working with a lower or higher transmit power than the quoted value, and the corresponding state can be formulated as follows:
\begin{equation}
  {\mathcal{L}_2}:{y_L(t)} = \sqrt {{\delta \cdot P_s}} s(t)+n(t), \delta \in (0,1)\cup (1,\infty). \hfill \\
\label{Eq_rogue1}
\end{equation}
The metric $\delta$ reflects the tendency of the IU and the seriousness of the illegitimate behaviors. $\delta>1$ indicates that the IU aims to enhance its own communication and bring the neighbors more interference, while $\delta<1$ indicates that the IU aims to occupy the channel over a large area and prevent other users of taking access to the channel in the area.

It is noted that in this case, only one user exists. If the user is working on the channel, it is either the LU or the IU during one slot. It is different from the case of illegitimate access, in which the LU and the IU are two entities. This difference will be apparently reflected in the modeling process below.

\subsection{Preliminary Spectrum Sensing Model: Composite Hypothesis Tests}
\subsubsection{Spectrum sensing model under illegitimate access}
Considering the probabilistic co-existence of the LU and the IU, the spectrum sensing problem under illegitimate access behaviors is formulated as a quaternary hypothesis test:
\begin{equation}
\begin{gathered}
  {\mathcal{S}_0}:y(t) = n(t), \hfill \\
  {\mathcal{S}_1}:y(t) = \sqrt {{P_s}} s(t) + n(t), \hfill \\
  {\mathcal{S}_2}:y(t) = \sqrt {{P_x}} x(t) + n(t), \hfill \\
  {\mathcal{S}_3}:y(t) = \sqrt {{P_s}} s(t) + \sqrt {{P_x}} x(t) + n(t), \hfill \\
\end{gathered}
\end{equation}
where ${\mathcal{S}_0}$, ${\mathcal{S}_1}$, ${\mathcal{S}_2}$, and ${\mathcal{S}_3}$ denote the states that no users exist, only the LU exists, only the IU exists, and the IU and the LU co-exist, respectively, and $y(t)$ is the observation of a certain sensor at the $t$-th slot.

Based on the signal characteristics, the difference among the hypotheses, in the view of the sensor, is the observation value's variance. Specifically, $y \sim \mathcal{N}(0,\sigma ^2)$, where the  parameter $\sigma^2$ is $\sigma _n^2$, $(P_s+\sigma _n^2)$, $(P_x+\sigma _n^2)$, and $(P_s+P_x+\sigma _n^2)$, under the hypotheses $\mathcal{S}_0$, $\mathcal{S}_1$, $\mathcal{S}_2$, and $\mathcal{S}_3$, respectively. Notably, $P_x$ is unknown and due to the superposition of the LU's signal and the IU's signal in ${\mathcal{S}_3}$, we have $\sigma^2  >\sigma_1^2$. Hence, the test can be rewritten as follows:
\begin{equation}
\begin{gathered}
  {\mathcal{S}_0}:\sigma^2  = \sigma_0^2, \hfill \\
  {\mathcal{S}_1}:\sigma^2  = \sigma_1^2,  \hfill \\
  {\mathcal{S}_2}:\sigma^2  >\sigma_0^2 ~\text{and}~ \sigma^2 \neq \sigma_1^2,  \hfill \\
  {\mathcal{S}_3}:\sigma^2  >\sigma_1^2,  \hfill \\
\end{gathered}
\label{Eq_quart}
\end{equation}
where $\sigma_0^2=\sigma_n^2$, $\sigma_1^2=P_s+\sigma_n^2$. Here, $\mathcal{S}_0$ and $\mathcal{S}_1$ are simple hypotheses whose parameter is a single value, while $\mathcal{S}_2$ and $\mathcal{S}_3$ are composite ones whose parameter's value space is a set.
In ${\mathcal{S}_2}$, $\sigma^2 \neq \sigma_1^2$ is set to distinguish ${\mathcal{S}_2}$ from ${\mathcal{S}_1}$. In fact, in $\mathcal{S}_2$, the value of $\sigma^2$ is continuous so that the probability of $\sigma^2= \sigma_1^2$ is zero, which means that it poses few effects on the results of the test.

Furthermore, let $p(P_x)$ denote the probability density function of the received power from the IU. Then, the probability distribution function $F({\sigma ^2})$  of the variance $\sigma^2$  is
\begin{equation}
F({\sigma ^2}) = \left\{ {\begin{array}{*{20}{l}}
  0,&{{\sigma ^2} < \sigma _0^2,} \\
  {a_{00}},&{{\sigma ^2} = \sigma _0^2,} \\
  {a_{00}+a_{01}\int_0^{{\sigma ^2} - \sigma _0^2} {p({P_x})d{P_x}},}&{\sigma _0^2 < {\sigma ^2} < \sigma _1^2,} \\
  a_{00} + a_{10}+a_{01}\int_0^{\sigma _1^2 - \sigma _0^2} {p({P_x})d{P_x}} ,&{{\sigma ^2} = \sigma _1^2,} \\
  {F(\sigma_1^2)+ a_{11}\int_0^{{\sigma ^2} - \sigma _1^2} {p({P_x})d{P_x}} },&{{\sigma ^2} > \sigma _1^2,}
\end{array}} \right.
\label{Eq_dens1}
\end{equation}
where $a_{ij}=P({\mathcal{L}_i},{\mathcal{I}_j})$.
Then, we have the probability density function of the observation value:
\begin{equation}
p(y_i) = \int_{\sigma _0^2}^{ + \infty } {\frac{1}{{\sqrt {2\pi {\sigma ^2}} }}} {e^{ - \frac{{{y_i^2}}}{{2{\sigma ^2}}}}}dF({\sigma ^2}),
\label{Eq_dens2}
\end{equation}
where $i \in \{1,2,...,N\}$.
\begin{figure}[t]
\centering
\includegraphics[width=1\linewidth]{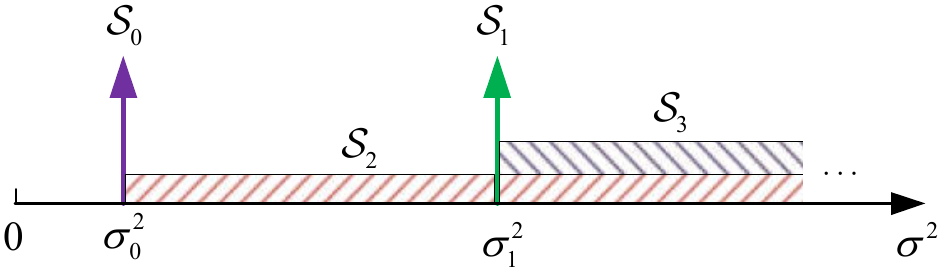}
\caption{Spectrum sensing under illegitimate access: A quaternary-hypothesis test.}
\label{Fig-four}
\end{figure}

\subsubsection{Spectrum sensing model under the rogue power control}
Based on (\ref{Eq_rogue0})(\ref{Eq_rogue1}), there are three channel states under the rogue power control, and correspondingly, the spectrum sensing problem can be modeled as a ternary hypothesis test:
\begin{equation}
\begin{gathered}
  {\mathcal{L}_0}:\sigma^2  = \sigma_0^2, \hfill \\
  {\mathcal{L}_1}:\sigma^2  = \sigma_1^2,  \hfill \\
  {\mathcal{L}_2}:\sigma^2  = \sigma_0^2+\delta (\sigma_1^2-\sigma_0^2), \delta \in (0,1)\cup (1,\infty).  \hfill \\
\end{gathered}
\end{equation}
Here, both the metric $\delta$ and its distribution are unknown, which means that the hypothesis ${\mathcal{L}_2}$ is composite.

\subsubsection{Summary}
The two cases are different in many aspects, such as external/internal attacks, one/two entities entity, and attack objectives. However, from the perspective of spectrum sensing, the two kinds of spectrum misuse behaviors show much similarity:
\begin{itemize}
  \item Both the tests consist of simple hypotheses without the IU and composite hypotheses with the IU.
  \item There are unknown parameters in both cases, i.e., the received power from the IU $P_x$ and the metric $\delta$, and though the parameters are different, they pose effects on spectrum sensing through the variances of the observations.
\end{itemize}
In the following, we will find that the two cases can be uniformly modeled through certain reasonable transformation.

\subsection{Uniform Spectrum Sensing Model: A Ternary Hypothesis Test}
Firstly, in the case of illegitimate access, though the quaternary hypothesis test in (\ref{Eq_quart}) gives the general formulation about the spectrum sensing problem under illegitimate access, the prior knowledge, such as the prior probabilities and the distribution of $P_x$, is hard to obtain and the equations (\ref{Eq_dens1})(\ref{Eq_dens2}) cannot be calculated. Furthermore, the hypothesis $\mathcal{S}_2$ with the variance $\sigma^2 >\sigma_0^2,~\sigma^2 \neq \sigma_1^2$, and the hypothesis $\mathcal{S}_3$ with the variance $\sigma^2 >\sigma_1^2$ are partly overlapping, just as shown in Fig. \ref{Fig-four}. As a result, it is hard to distinguish the two states.

To tackle the problems above, we transform the original quaternary hypothesis test about the illegitimate access to a ternary hypothesis test. The key rational is that our objective is to detect whether the channel is occupied and if it is occupied, recognize whether the IU exists, where the first part is to find out the spectrum holes, while the second part is to protect the scare resource of spectrum from being used illegitimately. To achieve the former part, $\mathcal{S}_0$ ought to be distinguished from the other hypotheses, while to achieve the latter one, the hypotheses $\mathcal{S}_0$ and $\mathcal{S}_1$ ought to be distinguish from the other two hypotheses $\mathcal{S}_2$ and $\mathcal{S}_3$. Hence, the objective can be achieved, even though the two hypotheses $\mathcal{S}_2$ and $\mathcal{S}_3$ are not distinguished. That is to say, we can combine $\mathcal{S}_2$ and $\mathcal{S}_3$ in one hypothesis, i.e., $\sigma^2  \in (\sigma _0^2,\sigma _1^2 ) \cup (\sigma _1^2 , + \infty )$.

Here, we find that the new hypothesis in the case of illegitimate access is identical to ${\mathcal{L}_2}$, i.e., the hypothesis of the IU launching rogue power control. Consequently, the two cases can be uniformly modeled as a ternary hypothesis test:
\begin{equation}
\left\{ \begin{gathered}
  {\mathcal{H}_0}:\sigma^2  = {\sigma _0^2}, \hfill \\
  {\mathcal{H}_1}:\sigma^2  = \sigma _1^2, \hfill \\
  {\mathcal{H}_2}:\sigma^2  \in (\sigma _0^2,\sigma _1^2 ) \cup (\sigma _1^2 , + \infty ), \hfill \\
\end{gathered}  \right.
\end{equation}
where the simple hypotheses $\mathcal{H}_0$ and $\mathcal{H}_1$ denote that the channel is idle and the channel is occupied by the LU, respectively, while the composite hypothesis $\mathcal{H}_2$ means that the IU exists, in which the IU can launch illegitimate access or rogue power control, and $\sigma_2^2$ is used to denote the unknown variance of the observation in $\mathcal{H}_2$.

\section{The Generalized Multi-Hypothesis Neyman-Pearson Criterion and the test rule}
In this section, based on the proposed ternary-hypothesis testing model, a novel test criterion called the GMNP criterion is proposed and the test rule is derived, which is decomposed into two subproblems:  a detection subproblem and a recognition subproblem. Then, the mutual effects between the two parts are analyzed. Finally, the upper bound of the detection performance is derived with the prior knowledge of the range of the unknown parameter.
\subsection{The Generalized Multi-Hypothesis Neyman-Pearson Criterion}
In the ternary hypothesis model, not only the prior probabilities are unknown, but also the prior distribution of $\sigma_2^2$ is hard to obtain. Moreover, it is conflicting to improve the detection probabilities of all hypotheses simultaneously. To tackle these issues, we build an optimization criterion named the generalized multi-hypothesis Neyman-Pearson (GMNP) criterion, where the probability of correctly identifying misuse behaviors, $\Pr ({\mathcal{H}_2}|{\mathcal{H}_2})$, is maximized, subjected to the constraints on the detection probabilities about $\mathcal{H}_0$ and $\mathcal{H}_1$ (i.e., $\Pr ({\mathcal{H}_0}|{\mathcal{H}_0})$ and $\Pr ({\mathcal{H}_1}|{\mathcal{H}_1})$), which is formulated as follows:
\begin{equation}
\begin{gathered}
  \mathop {\max }\limits_{{\mathcal{R}_0},{\mathcal{R}_1},{\mathcal{R}_2}} \Pr ({\mathcal{H}_2}|{\mathcal{H}_2}), \hfill \\
  s.t.\Pr ({\mathcal{H}_0}|{\mathcal{H}_0}) \geq \alpha ,\Pr ({\mathcal{H}_1}|{\mathcal{H}_1}) \geq \beta,  \hfill \\
\end{gathered}
\label{Eq_decision_region}
\end{equation}
where $\Pr(\mathcal{H}_i|\mathcal{H}_j),~i,j\in \{0,1\}$ denotes the probability of the decision $\mathcal{H}_i$ being made when the real state is $\mathcal{H}_j$, and $\alpha$ and $\beta$ are constants which play as the constraints on the basic test performance, $0.5<\alpha<1,~0.5<\beta<1$. In addition, ${\mathcal{R}_i}$ denotes the decision region of $\mathcal{H}_i$, that is,
\[\Pr ({\mathcal{H}_i}|{\mathcal{H}_j}) \triangleq \int_{{\mathcal{R}_i}} {p(\mathbf{y};{\mathcal{H}_j})} d\mathbf{y},\]
where $p(\mathbf{y};{\mathcal{H}_j})$ is the probability density function of the observation sequence $\mathbf{y}$ under ${\mathcal{H}_j}$.

Furthermore, we equivalently simplify the optimization criterion based on Lemma \ref{Lemma_equ}.

\begin{lem}
The optimization criterion in (\ref{Eq_decision_region}) is equivalent to the one below:
\begin{subequations}
\begin{align}
  \mathop {\max }\limits_{{\mathcal{R}_0},{\mathcal{R}_1},{\mathcal{R}_2}} \Pr ({\mathcal{H}_2}|{\mathcal{H}_2}), \hfill \\
  s.t.\Pr ({\mathcal{H}_0}|{\mathcal{H}_0}) = \alpha, \\
  \Pr ({\mathcal{H}_1}|{\mathcal{H}_1}) = \beta.
\end{align}
\label{Eq_lemma}
\end{subequations}
\label{Lemma_equ}
\end{lem}
\begin{rem}
In the traditional Neyman-Pearson criterion of the binary-hypothesis test, the detection probability about the alternative hypothesis is maximized with the constraint of the false alarm probability about the null hypothesis \cite{NP_tra}. Hence, the GMNP criterion in (\ref{Eq_decision_region}) and (\ref{Eq_lemma}) can be regarded as an extension of the Neyman-Pearson criterion in the generalized multi-hypothesis test.
Nevertheless, the main differences focus on one point: the multi-hypothesis test encounters the composite hypothesis. Multiple hypotheses increases the complexity of carving out the decision regions, while the composite hypothesis raises its difficulty. In the following, we focus on designing the test rule.
\end{rem}
\subsection{The Proposed Test Rule}
Firstly, to maximize the detection probability about spectrum misuse behaviors under the given detection probabilities of $\mathcal{H}_0$ and $\mathcal{H}_1$, we use the Lagrange multiplier method to construct the objective function:
\begin{equation}
\mathcal{F} =  \Pr({\mathcal{H}_2}|{\mathcal{H}_2}) + {\lambda _0}(\Pr({\mathcal{H}_0}|{\mathcal{H}_0}) - {\alpha}) + {\lambda _1}(\Pr({\mathcal{H}_1}|{\mathcal{H}_1}) - {\beta}).
\end{equation}
Further, we have
\begin{align}
  \mathcal{F} &= 1 - \int_{{\mathcal{R}_0}} {p(\textbf{y};{\mathcal{H}_2})} d\textbf{y} - \int_{{\mathcal{R}_1}} {p(\textbf{y};{\mathcal{H}_2})} d\textbf{y} \nonumber \\
  &~~~+ {\lambda _0}(\int_{{\mathcal{R}_0}} {p(\textbf{y};{\mathcal{H}_0})} d\textbf{y} - \alpha ) + {\lambda _1}(\int_{{\mathcal{R}_1}} {p(\textbf{y};{\mathcal{H}_1})} d\textbf{y} - \beta ) \nonumber \\
   &= \int_{{\mathcal{R}_0}} {[{\lambda _0}p(\textbf{y};{\mathcal{H}_0}) - p(\textbf{y};{\mathcal{H}_2})]} d\textbf{y} \nonumber \\
   &~~~+ \int_{{\mathcal{R}_1}} {[{\lambda _1}p(\textbf{y};{\mathcal{H}_1}) - p(\textbf{y};{\mathcal{H}_2})]} d\textbf{y} + 1 - {\lambda _0}\alpha  - {\lambda _1}\beta.
\label{Eq_optima}
\end{align}
In (\ref{Eq_optima}), we find that to maximize $\mathcal{F}$, in the decision region ${\mathcal{R}_0}$, ${\lambda _0}p(\textbf{y};{\mathcal{H}_0}) - p(\textbf{y};{\mathcal{H}_2}) > 0$, while in the decision region ${\mathcal{R}_1}$, ${\lambda _1}p(\textbf{y};{\mathcal{H}_1}) - p(\textbf{y};{\mathcal{H}_2}) > 0$, where the two constants $\lambda _0$ and $\lambda _1$ are decided to satisfy the constraints in ({\ref{Eq_lemma}).
Furthermore, to maximize $\Pr ({\mathcal{H}_2}|{\mathcal{H}_2})$, $\mathcal{H}_2$ is decided when
\begin{subequations}
\begin{align}
  \frac{{p({\mathbf{y}};{\mathcal{H}_2})}}{{p({\mathbf{y}};{\mathcal{H}_0})}} > {\lambda _0}, \\
  \frac{{p({\mathbf{y}};{\mathcal{H}_2})}}{{p({\mathbf{y}};{\mathcal{H}_1})}} > {\lambda _1}.
\end{align}
\label{Eq_lagran}
\end{subequations}

It is shown that to achieve the GMNP criterion in (\ref{Eq_decision_region}) and (\ref{Eq_lemma}), two likelihood ratios $\frac{{p({\mathbf{y}};{\mathcal{H}_2})}}{{p({\mathbf{y}};{\mathcal{H}_0})}}$ and $\frac{{p({\mathbf{y}};{\mathcal{H}_2})}}{{p({\mathbf{y}};{\mathcal{H}_1})}}$ are used in the detection process. Put (\ref{Eq_lemma}) and (\ref{Eq_lagran}) together, and we find that the constraint in (\ref{Eq_lemma}b) is related to the former likelihood ratio which includes $\mathcal{H}_0$ and $\mathcal{H}_2$, while the other constraint in (\ref{Eq_lemma}c) is related to the latter one which includes $\mathcal{H}_1$ and $\mathcal{H}_2$.
Hence, the optimization problem in (\ref{Eq_lemma}) can be decomposed into two subproblems, which are formulated as follows:
\begin{equation}
\begin{array}{l}
\mathop {\max }\limits_{{{\cal R}_0},{{\cal R}_2}} \Pr ({{\cal H}_2}|{{\cal H}_2}),\\
s.t.\Pr ({{\cal H}_0}|{{\cal H}_0}) = \alpha ,
\end{array}
\label{Eq_subpro1}
\end{equation}
and
\begin{equation}
\begin{array}{l}
\mathop {\max }\limits_{{{\cal R}_1},{{\mathcal R}_2}} \Pr ({{\cal H}_2}|{{\cal H}_2}),\\
s.t.\Pr ({{\cal H}_1}|{{\cal H}_1}) = \beta,
\end{array}
\label{Eq_subpro2}
\end{equation}
respectively, where the final solution of ${\mathcal R}_2$ is the intersection of the solutions of the two subproblems.
Hence, through the decomposition above, we simplify the former problem. More importantly, we find that the two subproblems have clear physical significance, based on which the two subproblems are named the detection subproblem and the recognition subproblem.

\subsubsection{Detection subproblem (Idle or busy)}
The first subproblem in (\ref{Eq_subpro1}) can be thought to stem from such a binary hypothesis test:
\begin{equation}
\begin{gathered}
  {\mathcal{H}_{0}}:{\sigma ^2} = \sigma _0^2, \hfill \\
  {\mathcal{H}_{2}}:{\sigma ^2} > \sigma _0^2,~\text{and}~{\sigma ^2} \neq \sigma _1^2. \hfill \\
\end{gathered}
\end{equation}

Let $\sigma _{2}$ denote the variance of the observation value under $\mathcal{H}_2$, and as $\sigma_2^2 > \sigma_0^2$, this problem is single-side. Then, to maximize $\Pr(\mathcal{H}_2|\mathcal{H}_2)$ while constraining $\Pr(\mathcal{H}_0|\mathcal{H}_0)$ to be $\alpha$, we exploit the likelihood ratio test (LRT):
\begin{align}
{L_0(\mathbf{y})}&=\frac{{p({\mathbf{y}};\mathcal{H}_2)}}{{p({\mathbf{y}};\mathcal{H}_0)}} = \frac{{\frac{1}{{{{(2\pi \sigma _{2}^2)}^{\frac{N}{2}}}}}\exp ( - \frac{{\sum\nolimits_{n = 0}^{N - 1} {y_n^2} }}{{2\sigma _{2}^2}})}}{{\frac{1}{{{{(2\pi \sigma _0^2)}^{\frac{N}{2}}}}}\exp ( - \frac{{\sum\nolimits_{n = 0}^{N - 1} {y_n^2} }}{{2\sigma _0^2}})}} \nonumber \\
&= {\left( {\frac{{\sigma _0^2}}{{\sigma _{2}^2}}} \right)^{\frac{N}{2}}}\exp \left(\frac{{\sigma _{2}^2 - \sigma _0^2}}{{2\sigma _0^2\sigma _{2}^2}}\sum\nolimits_{n = 0}^{N - 1} {y_n^2} \right) \mathop \gtreqless \limits_{\mathcal{H}_0}^{\mathcal{H}_2} \lambda_0.
\label{Eq_LRT}
\end{align}
As $\mathbf{y}$ works in the manner of $\sum\nolimits_{n = 0}^{N - 1} {y_n^2}$ in (\ref{Eq_LRT}), the test statistic is $Y = \sum\nolimits_{n = 0}^{N - 1} {y_n^2}$. $L_0(\mathbf{y})$ increases with $Y$, as $\sigma_{2}^2>\sigma_0^2$. Hence, to make $L_0(\mathbf{y})<\lambda_0$ means that $Y<\eta_0$, where $\eta_0$ is the corresponding threshold to $\lambda_0$. Further, under $\mathcal{H}_0$, $Y = \sum\nolimits_{n = 0}^{N - 1} {y_n^2}  \sim \chi _N^2(\sigma _0^2)$.
Hence, we have
\begin{equation}
\Pr(\mathcal{H}_0|\mathcal{H}_0) = \Pr (Y < {\eta_0}|{\mathcal{H}_0}) = 1-\frac{{\Gamma (\frac{N}{2},\frac{{{\eta_0}}}{{\sigma _0^2}})}}{{\Gamma (\frac{N}{2})}}=\alpha,
\label{Eq_dete}
\end{equation}
where $\Gamma(\cdot )$ and $\Gamma(\cdot,\cdot )$ are the Gamma function and upper incomplete Gamma function, respectively.
Hence, the decision threshold is derived as
\begin{equation}
\eta_0 = {\Gamma ^{ - 1}}\left(\frac{N}{2},\left( 1-\alpha \right)\Gamma (\frac{N}{2})\right)\sigma _0^2,
\label{Eq_eta0}
\end{equation}
where $\Gamma^{-1}(\cdot,\cdot)$ is the inverse incomplete Gamma function. Hence, the test rule for the detection subproblem is
\begin{equation}
Y\mathop \gtreqless \limits_{\mathcal{H}_0}^{\mathcal{H}_2}\eta_0.
\label{Eq_part1}
\end{equation}

\subsubsection{Recognition subproblem (Legitimate or illegitimate)}
The second subproblem (\ref{Eq_subpro2}) can be thought to stem from such a binary hypothesis test:
\begin{equation}
\begin{gathered}
  {\mathcal{H}_{1}}:{\sigma ^2} = \sigma_1^2, \hfill \\
  {\mathcal{H}_{2}}:{\sigma ^2} > \sigma _0^2,~\text{and}~{\sigma ^2} \neq \sigma _1^2. \hfill \\
\end{gathered}
\end{equation}

Different from the detection subproblem, the recognition subproblem is double-side, i.e., the unknown variance $\sigma_2^2$ may be either higher or lower than $\sigma_1^2$. Next, we apply the GLRT to tackle with the double-side problem. Then, asymptotic analysis is made about the detection performance. To bypass the complexity of obtaining the thresholds in the GLRT, another scheme, i.e., the Rao test, is also derived.

\paragraph{GLRT}

First, we calculate the unknown variance's maximum likelihood estimation (MLE) $\hat \sigma _2^2$ of $\sigma_2^2$:
\begin{equation}
\hat \sigma _2^2 = \arg \mathop {\max }\limits_{\sigma _2^2} p({\mathbf{y}};\sigma _2^2).
\end{equation}
Let $\frac{{\partial p(\textbf{y};{{\hat \sigma_2 }^2})}}{{\partial {{\hat \sigma_2 }^2}}} = 0$, and we have
\begin{equation}
{{\hat \sigma_2 }^2} = \frac{1}{N}\sum\nolimits_{n = 0}^{N - 1} {y_n^2}.
\end{equation}
Further, we have the likelihood ratio
\begin{align}
{L_1}&=\frac{{p({\mathbf{y}};{{\hat \sigma_2 }^2},\mathcal{H}_2)}}{{p({\mathbf{y}};\mathcal{H}_1)}}\nonumber \\
&={\left( {\frac{{N\sigma _1^2}}{{\sum\nolimits_{n = 0}^{N - 1} {y_n^2} }}} \right)^{\frac{N}{2}}}\exp \left(\frac{{\sum\nolimits_{n = 0}^{N - 1} {y_n^2} }}{{2\sigma _1^2}} - \frac{N}{2}\right)\mathop\gtreqless \limits_{\mathcal{H}_1}^{\mathcal{H}_2} \lambda_1.
\end{align}
As the test statistic $Y$ is $\sum\nolimits_{n = 0}^{N - 1} {y_n^2}$, ${L_1}$ is the function of $Y$ and we have
\begin{equation}\frac{{d{L_1}}}{{dY}} = \frac{1}{2}{\left( {N\frac{{\sigma _1^2}}{Y}} \right)^{\frac{N}{2}}}{e^{\frac{Y}{{2\sigma _1^2}}}}\frac{{Y - N\sigma _1^2}}{{Y\sigma _1^2}}.
\end{equation}
Therefore, when $Y>N{\sigma _1^2}$, $L_1$ increases with $Y$; otherwise, $L_1$ decreases with $Y$. Hence, the minimum of the likelihood ratio is achieved at $N{\sigma_1^2}$ and $\lambda_1>1$. Thus, $L_1<\lambda_1$ is equivalent to $\eta_1<Y<\eta_2$, where $\eta_1$ and $\eta_2$ are two solutions of $L_1(Y)=\lambda_1$, i.e.,
\begin{align}
{\left( {\frac{{N\sigma _1^2}}{{{\eta _1}}}} \right)^{\frac{N}{2}}}\exp (\frac{{{\eta _1}}}{{2\sigma _1^2}} - \frac{N}{2}) &= {\left( {\frac{{N\sigma _1^2}}{{{\eta _2}}}} \right)^{\frac{N}{2}}}\exp \left(\frac{{{\eta _2}}}{{2\sigma _1^2}} - \frac{N}{2}\right) \nonumber \\
&= {\lambda _1}.
\label{Eq_threshold1}
\end{align}
Simultaneously, when only the LU does exist, i.e., $\mathcal{H}_1$, $Y = \sum\nolimits_{n = 0}^{N - 1} {y_n^2}  \sim \chi _N^2(\sigma _1^2)$ and the condition below is satisfied:
\begin{align}
\Pr ({\mathcal{H}_1}|{\mathcal{H}_1}) &= \Pr ({\eta _1} < Y < {\eta _2}) = \frac{{\Gamma (\frac{N}{2},\frac{{{\eta _1}}}{{\sigma _1^2}}) - \Gamma (\frac{N}{2},\frac{{{\eta _2}}}{{\sigma _1^2}})}}{{\Gamma (\frac{N}{2})}} \nonumber \\
&=\beta.
\label{Eq_threshold2}
\end{align}
Hence, based on (\ref{Eq_threshold1}) and (\ref{Eq_threshold2}), the two thresholds $\eta_1$ and $\eta_2$ can be obtained. Finally, the test rule in this part is formulated as follows:
\begin{equation}
\left\{ {\begin{array}{*{20}{c}}
{{{\cal H}_1}~\text{is declared},}&{\text{if}~{\eta _1} < Y < {\eta _2},}\\
{{{\cal H}_2}~\text{is declared},}&{\text{otherwise.}}
\end{array}} \right.
\label{Eq_part2}
\end{equation}

\paragraph{Asymptotic analysis}

Considering the complexity of the GLRT, the thresholds are hard to be formulated in the closed-form manner. Nevertheless, when the number of samples is sufficiently large, some approximation can be made and the asymptotic performance (i.e., $N \rightarrow \infty $) is analyzed.

\begin{pro}
When $N$ is sufficiently large and the channel is occupied by the LU, we have $2\ln {L_1}\mathop  \sim \limits^a \chi _1^2$.
\end{pro}
Then, let $z^2=2\ln{L_1}$, where $z$ follows a Gaussian distribution with mean zero and variance 1, i.e., $z \sim \mathcal{N}(0,1)$, and we have
\begin{align}
\Pr ({\mathcal{H}_1}|{\mathcal{H}_1}) & \mathop  = \limits^a  \Pr (2\ln {L_1} < 2\ln {\lambda _1}|{\mathcal{H}_1}) \nonumber \\
&= \int_{ - \sqrt {2\ln {\lambda _1}} }^{\sqrt {2\ln {\lambda _1}} } {\frac{1}{{\sqrt {2\pi } }}} {\rho ^{\frac{{{z^2}}}{2}}}dz \nonumber \\
&= 2Q( - \sqrt {2\ln {\lambda _1}} ) - 1 = \beta,
\end{align}
where $Q(\cdot)$ is the complementary distribution function of the standard Gaussian. Hence, we can obtain the asymptotic solution of $\lambda_1$ as follows
\begin{equation}
{\lambda _1} \mathop  = \limits^a \exp \left[\frac{1}{2}{\left( {{Q^{ - 1}}(\frac{{1 + \beta }}{2})} \right)^2}\right].
\label{Eq_lambda1}
\end{equation}
In addition, based on (\ref{Eq_threshold1}), the approximate values of $\eta_1$ and $\eta_2$ are obtained.

Though the complexity of calculating the thresholds decreases, we still cannot obtain the closed-form expressions of the thresholds, which makes it not easy to evaluate the test performance. Nevertheless, we make another proposition below so that we can evaluate the asymptotic performance.
\begin{pro}
When $N$ is sufficiently large and the hypothesis $\mathcal{H}_2$ is true, $z^2=2\ln {L_1}(y)\mathop  \sim \limits^a {{\chi^{'}}_1^2(\theta  )}$, $z  \sim {\mathcal{N}(\sqrt \theta  ,1)}$, where ${{\chi^{'}}_1^2(\theta)}$ is a noncentral chi-square distribution with the noncentral parameter $\theta=(\sigma_2^2-\sigma_1^2)^2I(\sigma_1^2)$, and $I(\sigma_1^2)$ is the Fisher information. Specifically, the Fisher information is calculated below:
\begin{align}
  I(\sigma _1^2) &=  - E\left( {\frac{{{\partial ^2}\ln p(y;\sigma _1^2)}}{{{\partial ^2}\sigma _1^2}}} \right)\nonumber \\
  &=  - E\left( {\frac{\partial }{{\partial \sigma _1^2}}\left( {\frac{{\sum\nolimits_{i = 0}^{N - 1} {y_i^2} }}{{2\sigma _1^4}} - \frac{N}{{2\sigma _1^2}}} \right)} \right) \nonumber \\
   &=  - E\left( {\frac{N}{{2\sigma _1^4}} - \frac{{\sum\nolimits_{i = 0}^{N - 1} {y_i^2} }}{{\sigma _1^6}}} \right) = \frac{N}{{2\sigma _1^4}}.
\end{align}

\end{pro}
Then, we have $z  \sim {\mathcal{N}(\mu,1)},~\mu=\sqrt {(\frac{\sigma_2^2}{\sigma_1^2}-1)^2 \frac{N}{2}}$.
Hence, the asymptotic probability of $\mathcal{H}_2$ being wrongly detected as $\mathcal{H}_1$ is
\begin{align}
\Pr ({\mathcal{H}_1}|{\mathcal{H}_2}) & \mathop  = \limits^a \Pr ( -\sqrt {{2\ln {\lambda _1}} }  < z < \sqrt {{2\ln {\lambda _1}} }  |{\mathcal{H}_2}) \nonumber \\
&=\Pr (  {{Q^{ - 1}}(\frac{{1 + \beta }}{2})}  < z < {-{Q^{ - 1}}(\frac{{1 + \beta }}{2})}  |{\mathcal{H}_2}) \nonumber \\
&=\int_{  {{Q^{ - 1}}(\frac{{1 + \beta }}{2})}  }^{{-{Q^{ - 1}}(\frac{{1 + \beta }}{2})}  } {\frac{1}{\sqrt {2\pi }}{e^{ - \frac{{{\left(z -\mu \right)^2}}}{2}}}dz}.
\label{Eq_asympt}
\end{align}
Here, $2\ln{\lambda_1}=\left( {{Q^{ - 1}}(\frac{{1 + \beta }}{2})} \right)^2$ based on (\ref{Eq_lambda1}), and as $\beta<1$, $0.5<\frac{{1 + \beta }}{2}<1$, ${Q^{ - 1}}(\frac{{1 + \beta }}{2})<0$. So, $\sqrt {{2\ln {\lambda _1}}}=-{Q^{ - 1}}(\frac{{1 + \beta }}{2})$.

\paragraph{Rao test}
The Rao test has the same asymptotic performance with the GLRT, but it has a lower computation complexity, as the elements are easy to computed. So, we give the Rao test as follows:
\begin{align}
{T_R}(\mathbf{y})& = {\left. {\frac{{\partial \ln p(\mathbf{y};{\sigma ^2})}}{{\partial {\sigma ^2}}}} \right|_{{\sigma ^2}= \sigma _1^2}^2}{I^{ - 1}}(\sigma _1^2) \nonumber \\
&= {\left( {\frac{{\sum\nolimits_{n = 0}^{N - 1} {y_n^2}  - N\sigma _1^2}}{{2\sigma _1^4}}} \right)^2}\frac{{2\sigma _1^4}}{N} \nonumber \\
 &= \frac{{{{\left(\sum\nolimits_{n = 0}^{N - 1} {y_n^2}  - N\sigma _1^2\right)}^2}}}{{2N\sigma _1^4}}\mathop\gtreqless \limits_{\mathcal{H}_1}^{\mathcal{H}_2} \lambda_1^R.
\end{align}
Then, the two thresholds $\eta _1^R$ and $\eta _2^R$ about the test statistic $Y$ are formulated:
\begin{equation}
\left\{ \begin{array}{l}
\eta _1^R = (N - \sqrt {2N{\lambda _1^R}} )\sigma _1^2,\\
\eta _2^R = (N + \sqrt {2N{\lambda _1^R}} )\sigma _1^2.
\end{array} \right.
\label{Eq_rao}
\end{equation}
Simultaneously, we have
\begin{align}
\Pr ({\mathcal{H}_1}|{\mathcal{H}_1}) &= \Pr ({\eta _1^R} < Y < {\eta _2^R}) \nonumber \\
&= \frac{{\Gamma (\frac{N}{2},\frac{{{\eta _1^R}}}{{\sigma _1^2}}) - \Gamma (\frac{N}{2},\frac{{{\eta _2^R}}}{{\sigma _1^2}})}}{{\Gamma (\frac{N}{2})}}=\beta.
\label{Eq_threshold2}
\end{align}
Based on (\ref{Eq_rao}) and (\ref{Eq_threshold2}), the Rao test is built.

\begin{figure*}[!t]
\begin{align}
\Pr ({{\cal H}_2}|{{\cal H}_2}) & = 1 - \Pr ({\mathcal{H}_1}|{\mathcal{H}_2}) - \Pr ({\mathcal{H}_0}|{\mathcal{H}_2}) \nonumber \\
&\mathop = \limits^a 1- \left[Q \left({{Q^{ - 1}}(\frac{{1 + \beta }}{2})}-\mu \right)-Q \left(-{{Q^{ - 1}}(\frac{{1 + \beta }}{2})}-\mu\right)\right]-\left(1 - \frac{{\Gamma (\frac{N}{2},\frac{{{\eta _0}}}{{\sigma _2^2}})}}{{\Gamma (\frac{N}{2})}}\right) \nonumber \\
&=\frac{{\Gamma (\frac{N}{2},\frac{{{\eta _0}}}{{\sigma _2^2}})}}{{\Gamma (\frac{N}{2})}}-\left[Q \left({{Q^{ - 1}}(\frac{{1 + \beta }}{2})}-\mu \right)-Q \left(-{{Q^{ - 1}}(\frac{{1 + \beta }}{2})}-\mu\right)\right],
\label{Eq_asympt1}
\end{align}
\end{figure*}

\subsubsection{Mutual effects between the two subproblems}
When the GLRT is applied, based on (\ref{Eq_part1}) and (\ref{Eq_part2}), the decision regions are obtained about the test statistic $Y=\sum\nolimits_{n = 0}^{N - 1} {y_n^2}$:
\begin{equation}
{\mathcal{R}_i}:= \left\{ {\begin{array}{*{20}{c}}
  {Y<\eta_0},&{i=0,} \\
  {\eta_1<Y<\eta_2},&{i=1,} \\
  {\eta_0<Y<\eta_1, \text{or}~ Y>\eta_2}, &{i=2,}
\end{array}} \right.
\label{Eq_decision}
\end{equation}
where $\mathcal{R}_2$ is an intersection of the decision regions about $\mathcal{H}_2$ in the two subproblems (\ref{Eq_part1}) and (\ref{Eq_part2}) in order to satisfy the constraints. When the Rao test is applied, substitute $\eta _1^R$ and $\eta _2^R$ for $\eta _1$ and $\eta _2$ in (\ref{Eq_decision}), respectively, and we can obtain the corresponding decision regions.

In (\ref{Eq_decision}), one problem exists, i.e., $\eta_1$ may be lower than $\eta_0$. It means that there is one overlapping region between $\mathcal{R}_0$ and $\mathcal{R}_1$ and the constraints are not satisfied any more. It is true when the performance parameters $\alpha$ and $\beta$ satisfy a certain condition, just as shown in Theorem \ref{theorem_overlap} below.

\begin{thm}
If the GLRT is applied, the overlapping happens between the decision regions $\mathcal{R}_0$ and $\mathcal{R}_1$, when the parameters $\alpha$ and $\beta$ satisfy such a condition:
\begin{equation}
\beta  > \frac{{\Gamma (\frac{N}{2},\frac{{\eta _1^*}}{{\sigma _1^2}})}}{{\Gamma (\frac{N}{2})}} - \frac{{\Gamma (\frac{N}{2},\frac{{\eta _2^*}}{{\sigma _1^2}})}}{{\Gamma (\frac{N}{2})}},
\label{Eq_overlapping}
\end{equation}
where $\eta _1^{*} = {\Gamma ^{ - 1}}(\frac{N}{2},(1 - \alpha )\Gamma (\frac{N}{2}))\sigma _0^2$, $\eta _2^{*}$ is a solution to ${(\frac{{\eta _1^*}}{{\eta _2^*}})^{\frac{N}{2}}}\exp (\frac{{\eta _2^* - \eta _1^*}}{{2\sigma _1^2}}) = 1$, and $\eta_2^{*} \ne \eta_1^{*}$.
\label{theorem_overlap}
\end{thm}

When the condition is not satisfied in Theorem \ref{theorem_overlap}, i.e., there are no mutual effects between the two subproblems, based on (\ref{Eq_asympt}), we can obtain the asymptotic value of $\Pr ({{\cal H}_2}|{{\cal H}_2})$ just as formulated in (\ref{Eq_asympt1}),
where $\Pr ({\mathcal{H}_0}|{\mathcal{H}_2})$ is obtained based on (\ref{Eq_dete}). As $\mu>0$ and $\mu$ increases with the number of samples $N$, $\Pr ({\mathcal{H}_1}|{\mathcal{H}_2})$ decreases with $N$. Simultaneously, $\Pr ({\mathcal{H}_0}|{\mathcal{H}_2})$ decreases with $N$. Hence, as $N$ goes to infinity, $\forall \sigma _2^2 >\sigma _0^2,~\sigma _2^2 \ne \sigma _1^2,~\Pr ({\mathcal{H}_2}|{\mathcal{H}_2}) \to 1$.

On the other side, when the overlapping happens, we have to adjust the decision regions to avoid the overlapping and satisfy the constraints. Go back to (\ref{Eq_optima}), and to satisfy the constraints, the two thresholds $\lambda_0$ and $\lambda_1$ ought to be raised. However, as the overlapping has happened, the result of leveraging $\lambda_0$ is that $\eta_0$ increases in (\ref{Eq_LRT}) and the overlapping area is enlarged, which shows no help for satisfying the constraints, while both $\eta_1$ and $\eta_2$ increase with $\lambda_1$ in (\ref{Eq_threshold1}) and the decision region $\mathcal{R}_1$ is enlarged. So, when the condition is satisfied in Theorem \ref{theorem_overlap}, $\eta_1$ is set to be equal to $\eta_0$ so that the constraint $\Pr(\mathcal{H}_0|\mathcal{H}_0)=\alpha$ is satisfied, and to satisfy
\begin{equation}\Pr(\mathcal{H}_1|\mathcal{H}_1) = \frac{{\Gamma (\frac{N}{2},\frac{{{\eta _1}}}{{\sigma _1^2}})}}{{\Gamma (\frac{N}{2})}} - \frac{{\Gamma (\frac{N}{2},\frac{{{\eta _2}}}{{\sigma _1^2}})}}{{\Gamma (\frac{N}{2})}}={\beta},
\end{equation}
$\eta_2$ is reformulated as follows:
\begin{equation}
{\eta _2} = {\Gamma ^{ - 1}}\left(\frac{N}{2},\Gamma (\frac{N}{2},\frac{{{\eta _1}}}{{\sigma _1^2}}) - \Gamma (\frac{N}{2})\beta \right)\sigma _1^2.
\label{Eq_eta_2}
\end{equation}

As a result, the decision region is reformulated as follows:
\begin{equation}
{\mathcal{R}_i} := \left\{ {\begin{array}{*{20}{c}}
  {Y < {\eta _0}},&{{i=0,}} \\
  {{\eta _1} < Y < {\eta _2}},&{{i=1,}} \\
  {Y > {\eta _2}},&{{i=2,}}
\end{array}} \right.
\label{Eq_region2}
\end{equation}
where $\eta_0$ is calculated based on (\ref{Eq_eta0}), $\eta_1=\eta_0$, and $\eta_2$ is calculated based on (\ref{Eq_eta_2}).
\begin{thm}
If the Rao test is applied in the recognition part, the overlapping happens between the decision regions $\mathcal{R}_0$ and $\mathcal{R}_1$, when the parameters $\alpha$ and $\beta$ satisfy such a condition:
\begin{equation}
\beta  > \frac{{\Gamma (\frac{N}{2},\frac{{\eta _1^r}}{{\sigma _1^2}})}}{{\Gamma (\frac{N}{2})}} - \frac{{\Gamma (\frac{N}{2},\frac{{\eta _2^r}}{{\sigma _1^2}})}}{{\Gamma (\frac{N}{2})}},
\label{Eq_overrao}
\end{equation}
where $\eta _1^r={\Gamma ^{ - 1}}(\frac{N}{2},(1 - \alpha )\Gamma (\frac{N}{2}))\sigma _0^2$, and $\eta_2^r=2N\sigma_1^2-\eta_1^r$.
\label{Theorem_rao}
\end{thm}
When the overlapping happens, the analysis results are consistent with that when the GLRT is applied and the decision region is given in (\ref{Eq_region2}).
\begin{rem}
In conclusion, there are two cases (no overlapping happens/ the overlapping happens) in the process of designing the test rule and correspondingly, the decision regions are given in (\ref{Eq_decision}) and (\ref{Eq_region2}), respectively. It can be found in Theorem \ref{theorem_overlap} that the overlapping happens when the constraints are rigid, i.e., $\alpha$ and $\beta$ are high.  As a result, the decision region of $\mathcal{H}_2$ is compressed. Essentially, it is due to the multi-hypothesis characteristic of the test and there is a tradeoff between the optimization objective and the constraints, which will be analyzed in the following simulations (see Section V for details).
Similar to the proposed rule, the generalized maximized likelihood (ML) rule also works in cases without the prior knowledge and makes the detection relying on estimation of related parameters \cite{GML}.
One key difference between the generalized ML rule and the proposed rule is what detection criterion is applied. The generalized ML rule is to minimize the error probability, while the proposed rule aims to achieve the proposed GMNP criterion, in which the probability of correctly identifying misuse behaviors, $\Pr ({\mathcal{H}_2}|{\mathcal{H}_2})$, is maximized, subjected to the constraints on the detection probabilities about $\mathcal{H}_0$ and $\mathcal{H}_1$ (i.e., $\Pr ({\mathcal{H}_0}|{\mathcal{H}_0})$ and $\Pr ({\mathcal{H}_1}|{\mathcal{H}_1})$).
\end{rem}
\subsection{An Upper Bound of the Detection Performance: When the Range of the Unknown Variance is Known}
In this subsection, to further evaluate the performance of the proposed test rule, the prior knowledge about the range of the unknown variance $\sigma_2^2$ is given and the corresponding test rule is derived, which can be regard as an upper bound of the detection performance following the GMNP criterion. Here, the range of $\sigma_2^2$ means whether $\sigma_2^2$ is higher than $\sigma_1^2$, and when the range of $\sigma_2^2$ is known, the recognition problem turns from a double-side detection to a single-side detection. The test rule is given below.

\begin{thm}
When the range of $\sigma_2^2$, i.e., whether $\sigma_2^2>\sigma_1^2$ or $\sigma_0^2<\sigma_2^2<\sigma_1^2$ is true, is known, the decision region about the test statistic $Y=\sum\nolimits_{n = 0}^{N - 1} {y_n^2}$ is given as follows:

Case 1: $\sigma_0^2<\sigma_2^2<\sigma_1^2$
\begin{equation}
{\mathcal{R}_i} := \left\{ {\begin{array}{*{20}{c}}
  {Y < {\eta _0}},&{{i=0,}} \\
  {Y > {\eta _1}},&{{i=1,}} \\
  {{\eta _0} < Y < {\eta _1}},&{{i=2,}}
\end{array}} \right.
\end{equation}
where $\eta _0 = {\Gamma ^{ - 1}}(\frac{N}{2},(1-\alpha) \Gamma (\frac{N}{2}))\sigma _0^2$, and $\eta _1 = {\Gamma ^{ - 1}}(\frac{N}{2},\beta \Gamma (\frac{N}{2}))\sigma _1^2$.

Case 2: $\sigma_2^2>\sigma_1^2$
\begin{equation}
{\mathcal{R}_i} := \left\{ {\begin{array}{*{20}{c}}
  {Y < {\eta _0}},&{{i=0,}} \\
  {{\eta _0} < Y < {\eta _1}},&{{i=1,}} \\
  {Y > {\eta _1}},&{{i=2,}}
\end{array}} \right.
\end{equation}
where $\eta _0 = {\Gamma ^{ - 1}}(\frac{N}{2},(1-\alpha) \Gamma (\frac{N}{2}))\sigma _0^2$, and $\eta_1={\Gamma ^{ - 1}}(\frac{N}{2},\Gamma (\frac{N}{2},\frac{{{\eta _0}}}{{\sigma _1^2}}) - \beta \Gamma (\frac{N}{2}))\sigma_1^2$.
\label{Theorem_upper}
\end{thm}

\section{Cooperative Spectrum Sensing Under Spectrum Misuse Behaviors}
In this section, to further improve the detection performance, cooperative spectrum sensing \cite{CSS_zhangwei,Optimal_CSS} is conducted based on a distributed detection framework, where there are $K$ sensors and a FC. To decrease the communication cost, sensors report local decisions to the FC in which data fusion-based global decision is made. We consider that the reporting channel is ideal and the decisions are correctly received by the FC.
Here, let $r_k$ and $\mathbf{r}$ denote the report of the $k$-th sensor, $r_k \in \{\mathcal{H}_0,\mathcal{H}_1,\mathcal{H}_2\}$, and  the whole reports, i.e., $\mathbf{r}=[r_1,r_2,...,r_K]$, respectively, where sensors' reports are considered to be conditionally independent.

Next, the FC makes data fusion-based global decision, denoted as $\mathcal{D}$. First of all, the global GMNP criterion is built to maximize the global probability of the IU being correctly detected under the constraints about the detection probabilities of $\mathcal{H}_0$ and $\mathcal{H}_1$:
\begin{equation}
\begin{gathered}
  \max \Pr ({\mathcal{D}=}{\mathcal{H}_2}|{\mathcal{H}_2}), \hfill \\
  s.t.\;\Pr ({\mathcal{D}={\mathcal{H}_0}}|{\mathcal{H}_0}) \geqslant {\alpha _f},\Pr ({\mathcal{D}={\mathcal{H}_1}}|{\mathcal{H}_1}) \geqslant {\beta _f}, \hfill \\
\end{gathered}
\label{Eq_global1}
\end{equation}
where  $\alpha <{\alpha _f}<1, \alpha<{\beta_f}<1$. Then, the problem is what global decision is made when certain combination of reports $\mathbf{r}$ is received.

\begin{pro} When the distances between sensors are small enough compared with the distances from the sensors to the IU and LU, it is reasonable to consider that all sensors have identical sensing performance. Then, the reports $\mathbf{r}$ can be simply denoted as $\mathbf{d}=(d_0,d_1,d_2)$, where $d_i$ is the number of sensors who claim $\mathcal{H}_i$, $i \in \{1,2,3\}$. Here, $d_0+d_1+d_2=K$, and the total number of possible $\mathbf{d}$ is $L=\left( {\begin{array}{*{20}{c}}
{K + 2}\\
2
\end{array}} \right) = \frac{{(K + 2)(K + 1)}}{2}$. Then, we have
\end{pro}
\begin{align}
  \Pr (\mathbf{d}|{\mathcal{H}_i}) &= \left( \begin{array}{*{20}{c}}
  K  \\
  {d_0}
\end{array}  \right)\Pr {({\mathcal{H}_0}|{\mathcal{H}_i})^{{d_0}}} \times \left( \begin{array}{*{20}{c}}
  K - {d_0} \\
  {d_1}
\end{array}  \right)\nonumber \\
&~~~ \Pr {({\mathcal{H}_1}|{\mathcal{H}_i})^{{d_1}}}\times \left( \begin{array}{*{20}{c}}
  K - {d_0} - {d_1} \\
  {d_2}
\end{array}  \right)\Pr {({\mathcal{H}_2}|{\mathcal{H}_i})^{{d_2}}} \nonumber \\
   &= \frac{{K!}}{{\prod\limits_{n = 0}^2 {{d_n}!} }}\prod\limits_{j = 0}^2 {\Pr {{({\mathcal{H}_j}|{\mathcal{H}_i})}^{{d_j}}}}.
\label{Eq_fusion}
\end{align}

The global decision is to make choices among $\mathcal{H}_0$, $\mathcal{H}_1$, and $\mathcal{H}_2$ for a certain $\mathbf{d}$. Let $S_i$, denote the set of $\mathbf{d}$ in which the decision $\mathcal{H}_i$ is made, $i \in \{1,2,3\}$, where the sets $S_0, S_1, S_2$ are complete and mutually exclusive. Therefore, from the perspective of optimizing the global decision, the global criterion in (\ref{Eq_global1}) can be rewritten as follows:
\begin{equation}
\begin{gathered}
  \mathop {\max }\limits_{{S_0},{S_1},{S_2}} \sum\limits_{\mathbf{d} \in {S_2}} {\Pr (\mathbf{d}|{\mathcal{H}_2})},  \hfill \\
  s.t.\sum\limits_{\mathbf{d} \in {S_0}} {\Pr (\mathbf{d}|{\mathcal{H}_0})}  \geqslant {\alpha _f},\sum\limits_{\mathbf{d} \in {S_1}} {\Pr (\mathbf{d}|{\mathcal{H}_1})}  \geqslant {\beta _f}. \hfill \\
\end{gathered}
\label{Eq_global2}
\end{equation}

Further, as ${\Pr (\mathcal{H}_i|{\mathcal{H}_j})}$, $i \in \{0,1,2\}, j\in\{0,1\}$ is known based on the local thresholds, ${\Pr (\mathbf{d}|{\mathcal{H}_j})}, j\in\{0,1\}$ can be calculated based on (\ref{Eq_fusion}), which means that we can find the sets who satisfy the constraints. Nevertheless, as ${\Pr (\mathcal{H}_i|{\mathcal{H}_2})}$ is unknown, ${\Pr (\mathbf{d}|{\mathcal{H}_2})}$ cannot be obtained. That is, we cannot directly pick out the optimal solution $(S_0^{*},S_1^{*},S_2^{*})$ among the ones who satisfy the constraints. Further, we find that the number of elements in the decision region $\mathcal{S}_2$ can reflect the optimization objective in Eq. (\ref{Eq_global1}), i.e., ${\Pr (\mathcal{D}=\mathcal{H}_2|{\mathcal{H}_2})}$, which is based on two points:
\begin{itemize}
  \item As the number of possible $\mathbf{d}$ is $L=\frac{{(K + 2)(K + 1)}}{2}$, where $K$ is the number of sensors, the value space is separated into $L$ pieces and the decision regions $\mathcal{S}_0$, $\mathcal{S}_1$, and $\mathcal{S}_2$ consist of certain numbers of pieces. Clearly, $L$ increases quadratically with $K$ and when $L$ is large enough, the number of pieces in the decision region, to some degree, represents the corresponding detection probability.
  \item In general, for certain $\mathbf{d}$, the higher the probabilities $\Pr(\mathbf{d}|\mathcal{H}_0)$ and $\Pr(\mathbf{d}|\mathcal{H}_1)$ are, the lower $\Pr(\mathbf{d}|\mathcal{H}_2)$ is, and vice versa. In the process of satisfying the constraints, through maximizing the number of elements in $\mathcal{S}_2$, $\mathbf{d}$ with high $\Pr(\mathbf{d}|\mathcal{H}_2)$ is left behind for $\mathcal{S}_2$, which is consistent with the optimization objective.
\end{itemize}

Hence, we prefer to choose the solution in which $S_2$ has more elements, in order to maximize the detection probability $\Pr ({\mathcal{H}_2}|{\mathcal{H}_2})$ based on (\ref{Eq_global2}).
Hence, the optimization problem in (\ref{Eq_global2}) is approximately transformed into
\begin{equation}
\begin{gathered}
  \mathop {\max }\limits_{{S_0},{S_1}} |{S_2}|, \hfill \\
  s.t.\;\sum\limits_{{\mathbf{d}} \in {S_0}} {\Pr ({\mathbf{d}}|{\mathcal{H}_0})} \geqslant {\alpha _f}, \hfill \\
  ~~~~~\sum\limits_{{\mathbf{d}} \in {S_1}} {\Pr ({\mathbf{d}}|{\mathcal{H}_1})} \geqslant {\beta _f}, \hfill \\
  ~~~~~|{S_0}| + |{S_1}| + |{S_2}| = \frac{{(K + 2)(K + 1)}}{2}, \hfill \\
\end{gathered}
\label{Eq_set}
\end{equation}
where $|S_i|$ denotes the number of elements in the set $S_i$.

Number all the possible combinations $\mathbf{d}$, and use $\mathbf{d}^{(i)}$ to denote the $i$th one, $i \in \{1,2,...,L\}$. Let $Y$  denote a $L \times 3$ array, where if $\mathbf{d}^{(i)}$ is included in the set $\mathcal{S}_j$, $y_{ij}=1$ and otherwise, $y_{ij}=0$. Then, the optimization problem in (\ref{Eq_set}) can be reformulated equivalently as follows:
\begin{equation}
\begin{gathered}
  \mathop {\min }\limits_{Y} ~\sum\limits_{i=1}^L {(y_{i1}+y_{i2})},  \hfill \\
  s.t.~~\sum\limits_{i=1}^L {y_{i1}\cdot c_{i1}}\geq {\alpha_f} ,\hfill \\
  ~~~~~~\sum\limits_{i=1}^L {y_{i2}\cdot c_{i2}}\geq {\beta_f}, \hfill \\
  ~~~~~~y_{i1}+y_{i2} \in \{0,1\},i=1,2,...,L, \hfill \\
  ~~~~~~y_{ij} \in \{0,1\}, i=1,2,...,L,j=1,2,\hfill \\
\end{gathered}
\label{Eq_NP}
\end{equation}
where $c_{i1}$ and $c_{i2}$ denote $\Pr (\mathbf{d}^{(i)}|{\mathcal{H}_0})$ and $\Pr (\mathbf{d}^{(i)}|{\mathcal{H}_1})$, respectively.
This problem is that of integer linear programming (ILP), while the ILP is NP-hard. Hence, the challenge is how to effectively make the assignment about $\mathbf{d}$  between $\mathcal{H}_0$ and $\mathcal{H}_1$ in order to satisfy the constraints.

Intuitively, to satisfy the constraints while maximizing $|{S_2}|$, $\mathbf{d}$ with high probabilities $\Pr(\mathbf{d}|\mathcal{H}_0)$ and $\Pr(\mathbf{d}|\mathcal{H}_1)$ are assigned to $S_0$ and $S_1$, respectively. Nevertheless, the two processes cannot be done independently, as there are potential collisions between them, i.e., there may be some $\mathbf{d}$ belonging to both $S_0$ and $S_1$. Therefore, we set a rule, in which $\mathbf{d}$ is assigned to the hypothesis with a higher probability when the collision happens, and give Algorithm 1. Specifically, in the beginning, $S_0$, $S_1$, and $S_2$ are empty, use $\Omega$ to denote the whole set of $\mathbf{d}$, and $U=\Omega-{S_0}\bigcup {S_1}$. Then, independently, $\mathbf{d}$ in $U$ is assigned to $S_0$ and $S_1$ in the descending orders of $\Pr(\mathbf{d}|{\mathcal{H}_0})$ and $\Pr(\mathbf{d}|{\mathcal{H}_0})$, respectively, until the constraints are satisfied (Line 11-18 in Algorithm 1). Let $X={S_0} \cap {S_1}$ and use $\mathbf{x}^{(i)}$ to denote the $i$th element in $X$. If $X\neq \emptyset$, compare $\Pr(\mathbf{x}^{(i)}|\mathcal{H}_0)$ and $\Pr(\mathbf{x}^{(i)}|\mathcal{H}_1)$, $\mathbf{x}^{(i)}$ is abandoned by the set with a lower probabilities (Line 20-28 in Algorithm 1), and then return to Line 10 to continue until $X$ is empty. When the iteration ends, $S_2=\Omega-{S_0}\bigcup {S_1}$.

Finally, it is pointed out that the algorithm's optimality is closely related to the constraints and when the constraints, i.e., $\alpha_f$ and $\beta_f$, are not too high, the algorithm is optimal for (\ref{Eq_NP}). Specifically, in Algorithm 1, we find that when the constraints are not too high, no collision happens so that all the elements in $S_0$ and $S_1$ are obtained based on the descending orders of $\Pr(\mathbf{d}|{\mathcal{H}_0})$ and $\Pr(\mathbf{d}|{\mathcal{H}_1})$, respectively, and as a result, $|S_0|+|S_1|$ obtains the smallest value so that the optimal $S_2$ is achieved. However, as the constraints increase, collisions increase and the performance may decrease.

\begin{algorithm}[t]
\caption{The Cooperative Spectrum Sensing under Spectrum Misuse Behaviors}
\begin{algorithmic}[1]
\STATE \textbf{//Local decision}\\
\IF {The condition (\ref{Eq_overlapping}) about $\alpha$ and $\beta$ is satisfied, i.e., the overlapping happens,}
    \STATE Calculate $\eta_0$ and $\eta_2$ based on (\ref{Eq_eta0}) and (\ref{Eq_eta_2}), respectively, $\eta_1 \leftarrow \eta_0$, and make local decisions based on (\ref{Eq_region2}).\\
\ELSE
    \STATE Calculate the thresholds based on (\ref{Eq_eta0}), (\ref{Eq_threshold1}), and (\ref{Eq_threshold2}) and make decisions based on (\ref{Eq_decision}).\\
\ENDIF
\STATE \textbf{//Data fusion and global decision}
\STATE $S_0\leftarrow \emptyset$, $S_1\leftarrow \emptyset$, $S_2\leftarrow \emptyset$, $\Omega\leftarrow \{\mathbf{d}^{(1)}, \mathbf{d}^{(2)},...,\mathbf{d}^{(L)}\}$, $X \leftarrow\Omega$
\WHILE {$X \neq\emptyset$}
\STATE $U=\Omega-{S_0}\bigcup {S_1}$, $U_0 \leftarrow U$, $U_1 \leftarrow U$
\WHILE {$\sum\limits_{\mathbf{d}\in S_0} {\Pr ({\mathbf{d}}|{{\cal H}_0})}<\alpha_f$}
\STATE $\mathbf{d}^{*}  \leftarrow \arg \mathop {\max }\limits_{\mathbf{d}\in {U_0}} \Pr(\mathbf{d}|{\mathcal{H}_0})$
\STATE ${S_0} \leftarrow {S_0} \cup \mathbf{d}$, $U_0\leftarrow {U_0} - \mathbf{d}$
\ENDWHILE
\WHILE {$\sum\limits_{\mathbf{d}\in S_1} {\Pr ({\mathbf{d}}|{{\cal H}_1})}<\beta_f$}
\STATE $\mathbf{d}^{*}  \leftarrow \arg \mathop {\max }\limits_{\mathbf{d}\in {U_1}} \Pr(\mathbf{d}|{\mathcal{H}_1})$
\STATE ${S_1} \leftarrow {S_1} \cup \mathbf{d}$, $U_1\leftarrow {U_1} - \mathbf{d}$
\ENDWHILE
\STATE $X\leftarrow {S_0}\bigcap {S_1}$, $N_x\leftarrow |X|$
\IF {$N_x\neq 0$}
\FOR {$i=1,...,N_x$}
\IF {$\Pr(\mathbf{x}^{(i)}|\mathcal{H}_0)>\Pr(\mathbf{x}^{(i)}|\mathcal{H}_1)$}
\STATE $S_1\leftarrow S_1-\mathbf{x}^{(i)}$
\ELSE
\STATE $S_0\leftarrow S_0-\mathbf{x}^{(i)}$
\ENDIF
\ENDFOR
\ENDIF
\ENDWHILE
\STATE $S_2 \leftarrow \Omega-S_0-S_1$.
\end{algorithmic}
\end{algorithm}

\section{Simulations}
\subsection{Basic Simulation Setup}
In the following, we consider an IEEE 802.22 simulation environment \cite{ieee802_22}. An LU (e.g., an advanced base station) is located at the origin (0 m, 0 m) and $K$ sensors are randomly distributed in a 100 m $\times$ 100 m square area with the center at (10 km, 0 m). The bandwidth of the spectrum band is 6 MHz.  The IU (e.g., an illegitimate broadcast station) is located randomly and it is over 1000 m away from the sensors. Without special statement, the number of samples is $N=300$, which corresponds to the sampling time 0.05 ms, and the number of sensors is $K=20$. The noise variance is $\sigma_n^2=10^{-5}$ Watt. The ratio of the received power from the LU $P_s$ to the noise variance at (10 km, 0 m) is set as -5dB. Sensors make ternary decisions and report the results to the FC, in which reports are fused and global decisions are made based on the proposed algorithm.

\subsection{Sensing with a Single Sensor}
To evaluate the performance with a single sensor, four schemes are included in the following simulations:
\begin{itemize}
  \item The scheme given in Theorem \ref{Theorem_upper}, named \emph{Upper bound}, where the information on the prior range of the variance $\sigma_2^2$ is exploited in the test.
  \item The proposed scheme named \emph{P-GLRT}, where the GLRT is exploited in the recognition part.
  \item The proposed scheme named \emph{P-Rao}, where the Rao test is exploited in the recognition part.
  \item The scheme named \emph{Asymptotic performance} given in (\ref{Eq_asympt1}), which is obtained through asymptotic analysis about sensing performance with a single sensor of both P-GLRT and P-Rao.
\end{itemize}

\begin{figure}[t]
\centering
\includegraphics[width=1\linewidth]{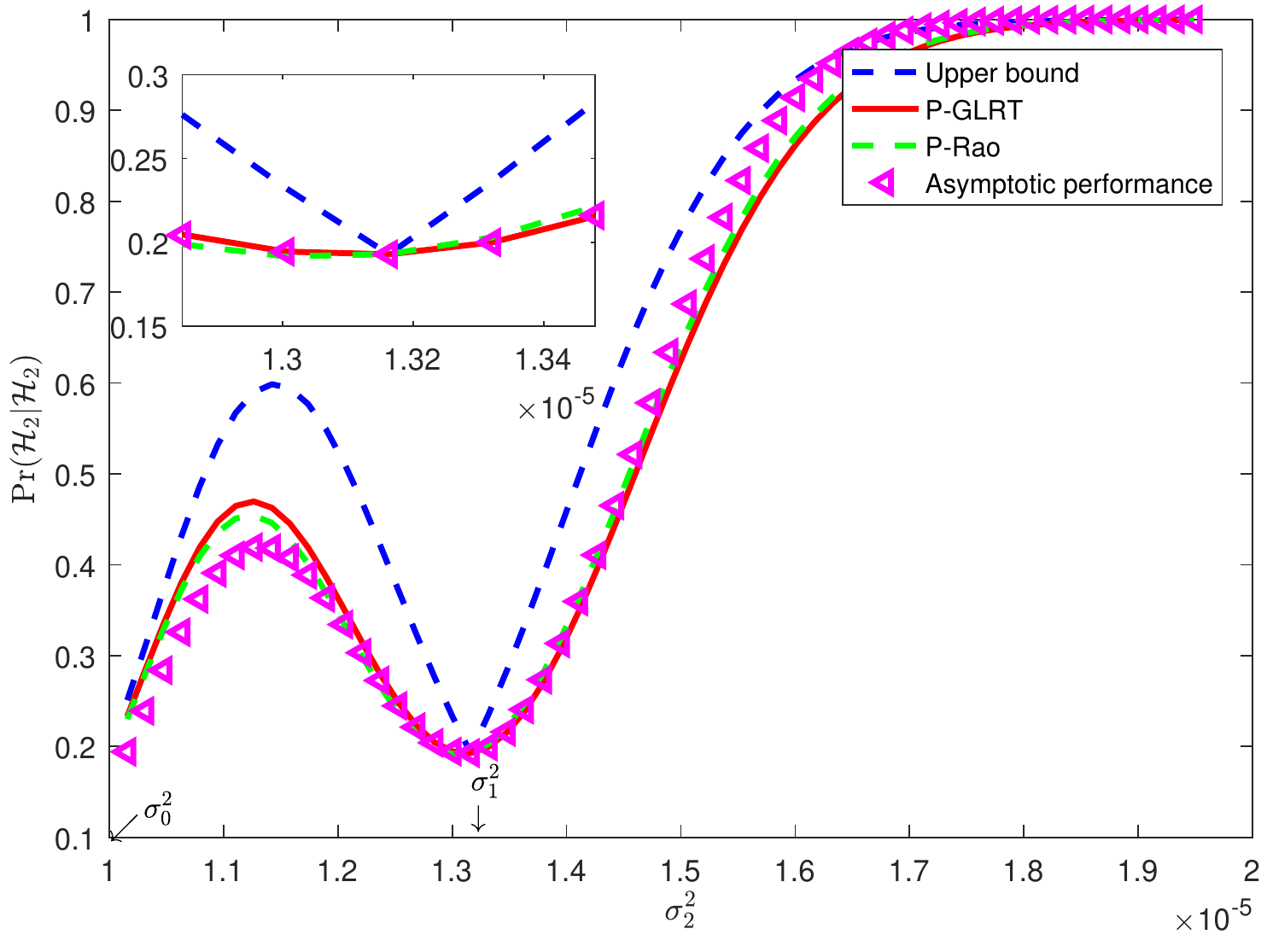}
\caption{The probability of a single sensor correctly detecting the existence of the IU versus the unknown variance $\sigma_2^2$. $\alpha$ and $\beta$ are set to 0.8.}
\label{Fig-iu-local}
\end{figure}
Fig. \ref{Fig-iu-local} illustrates the performance of a single sensor correctly detecting the existence of the IU versus the unknown parameter $\sigma_2^2$.
With the assistance of the prior knowledge about the range of the variance $\sigma_2^2$, the double-side detection is simplified as a single-side one and the upper bound of the detection performance is obtained. Hence, it is observed that the upper bound shows apparently higher performance than the other schemes. The two proposed schemes, P-GLRT and P-Rao, show similar performances and the asymptotic performance is close to these of the two proposed schemes, as the number of samples $N$ is 300, which is sufficiently large.
More importantly, when $\sigma_2^2$ approaches either $\sigma_0^2$ or $\sigma_1^2$, the detection performances of all schemes deteriorate and the gaps between the schemes are shorten.
That is, it becomes hard to distinguish $\mathcal{H}_2$ from either $\mathcal{H}_0$ or $\mathcal{H}_1$, which is due to the high similarity between the hypotheses from the perspective of testing, which can be seen as the worst cases that are determined by the constraints $\alpha$ and $\beta$. Hence, from the perspective of properly improving the performance in the worst cases, the constraints $\alpha$ and $\beta$ cannot be set too high.
On the other side, when the distances from $\sigma_2^2$ to $\sigma_0^2$ and $\sigma_1^2$ increase, the performances get better and in particular, when $\sigma_2^2$ is over $\sigma_1^2$, the gaps gradually diminish.

\begin{figure}[t]
\centering
\includegraphics[width=1\linewidth]{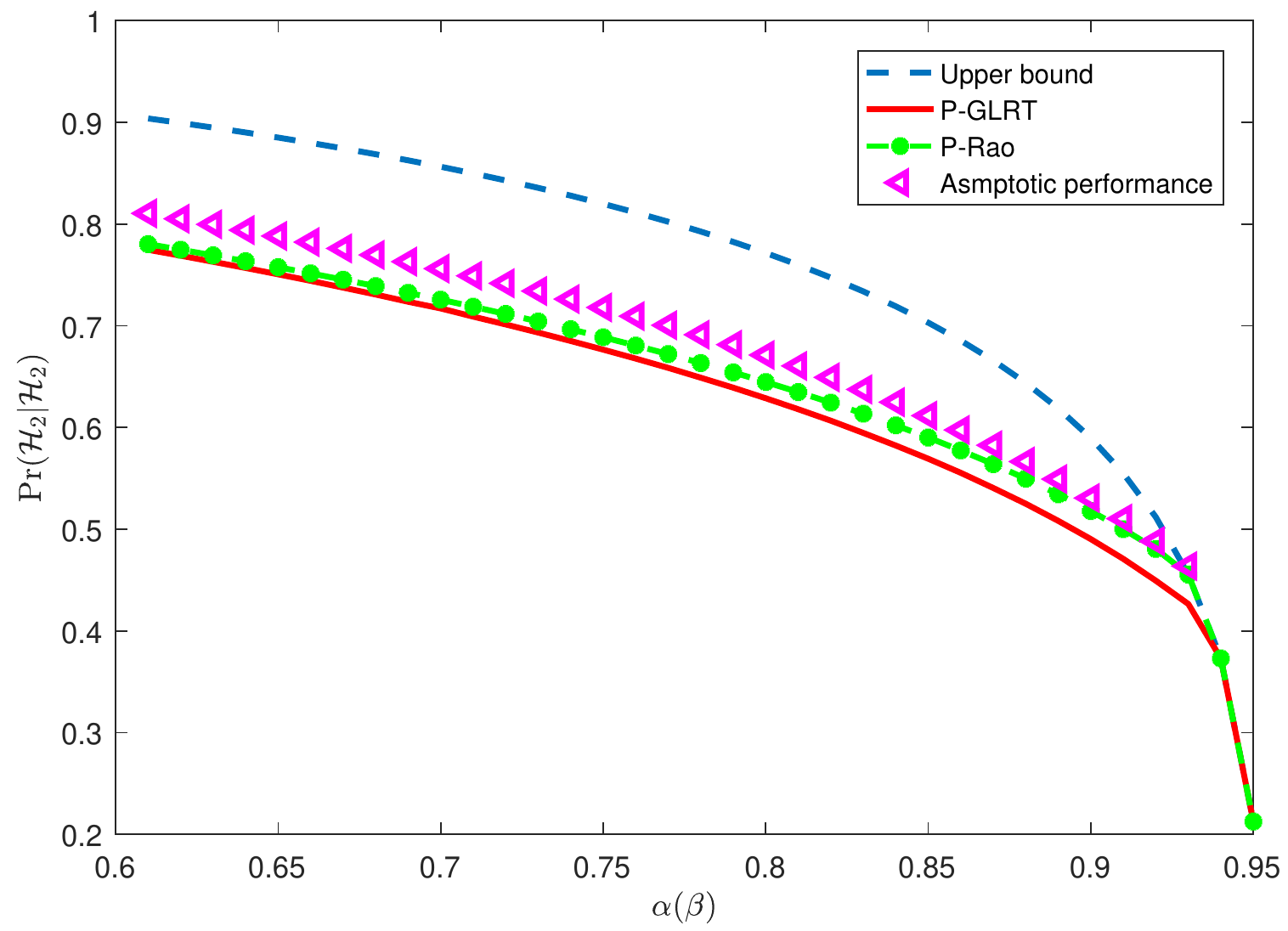}
\caption{Receiver operating characteristic (ROC) curves: $\Pr(\mathcal{H}_2|\mathcal{H}_2)$ versus $\Pr(\mathcal{H}_0|\mathcal{H}_0)$ and $\Pr(\mathcal{H}_1|\mathcal{H}_1)$, i.e., $\alpha$ and $\beta$. Here, $\alpha=\beta$, $N=300$, and the ratio of the received power from the IU $P_x$ to the noise variance is -3dB.}
\label{Fig-roc-local}
\end{figure}

\begin{figure}[!t]
\centering
\includegraphics[width=1\linewidth]{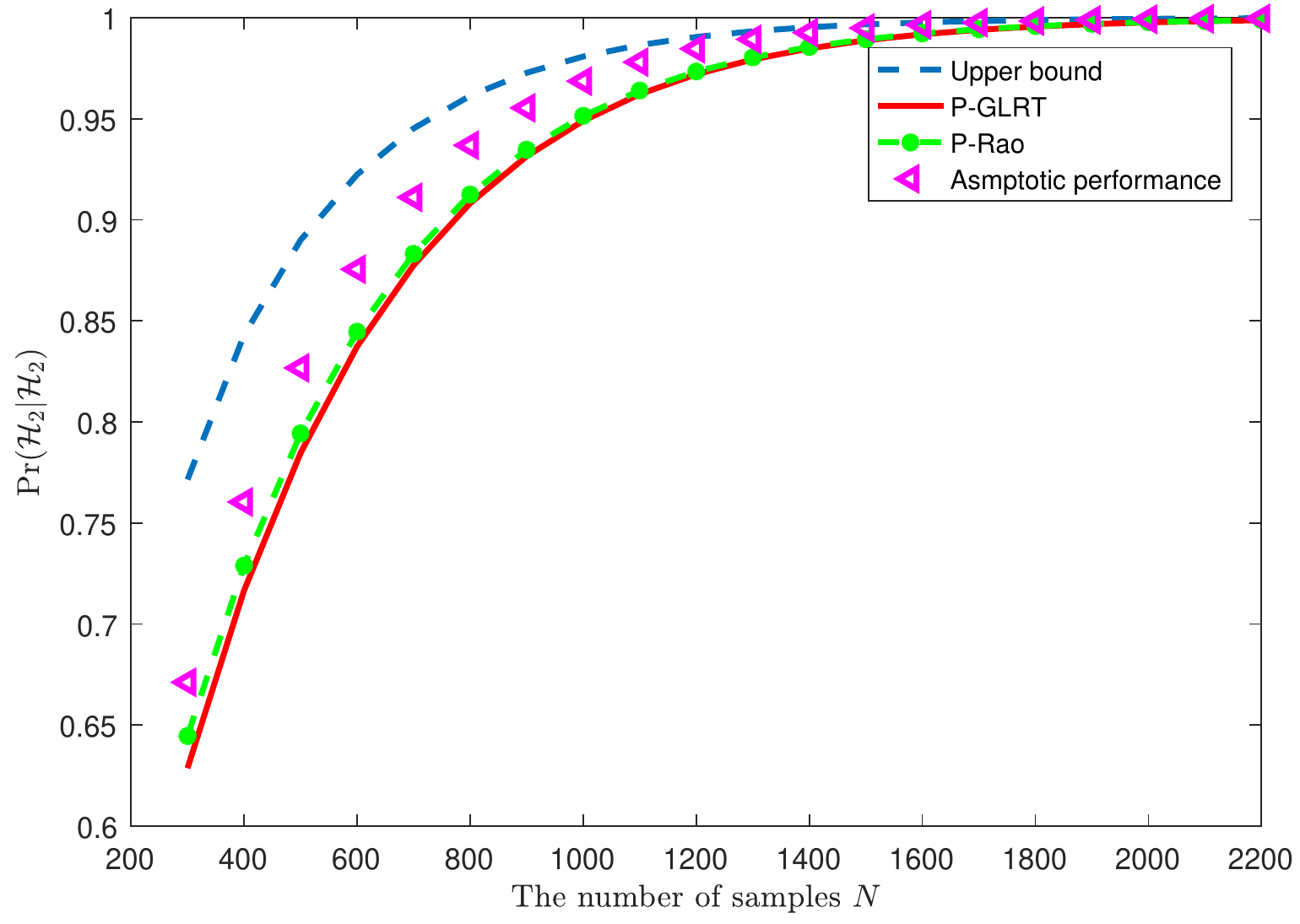}
\caption{The detection probability $\Pr(\mathcal{H}_2|\mathcal{H}_2)$ versus the number of samples $N$. Here $\alpha=\beta=0.8$, and the ratio of the received power from the IU $P_x$ to the noise variance is -3dB.}
\label{Fig-sample-local}
\end{figure}
As shown in Fig. \ref{Fig-roc-local}, when the requirements about $\Pr(\mathcal{H}_0|\mathcal{H}_0)$ and $\Pr(\mathcal{H}_1|\mathcal{H}_1)$ improve, i.e., $\alpha,~\beta$ increase, the probability of the IU being correctly detected decreases. It is implied that a tradeoff has to be achieved when tackling with detection of both the IU and the LU, which is similar to the tradeoff between the false-alarm probability and the detection probability in the traditional binary hypothesis test. In addition, when $\alpha=\beta>0.92$, the mutual effects between the detection part and the recognition part exist and the asymptotic performance cannot be obtained so that its last three points isn't curved.

Fig. \ref{Fig-sample-local} shows the detection performances increases with the number of samples $N$ and the gaps gradually diminish. In particular, when $N$ increases from 300 to 600, the proposed schemes' detection probabilities improve by nearly 33 percent.

\subsection{Sensing with Multiple Sensors}

\begin{figure}[!t]
\centering
\includegraphics[width=1\linewidth]{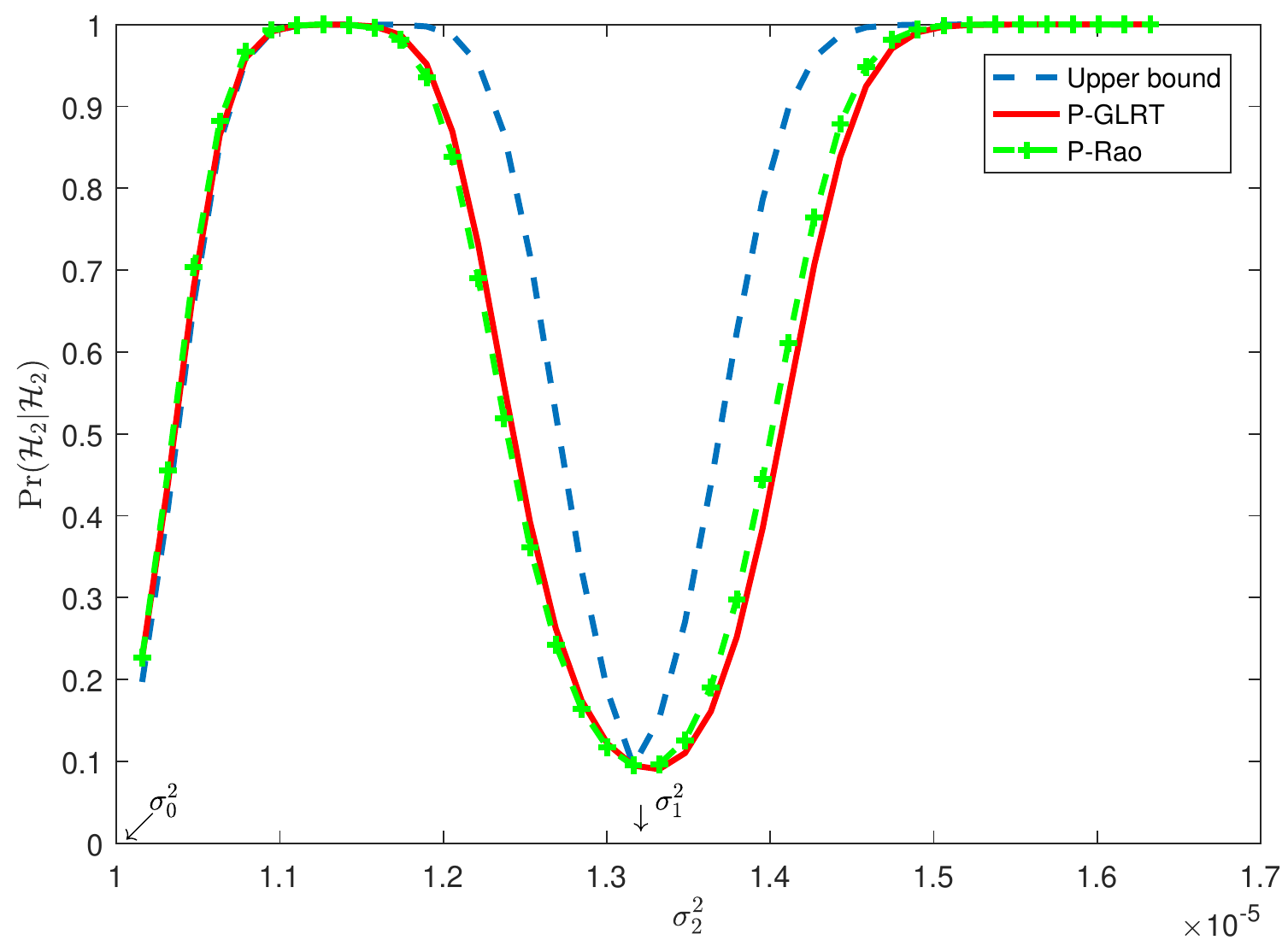}
\caption{The global performance of correctly detecting the existence of the IU versus the unknown parameter $\sigma_2^2$. Here, the number of samples $N$ is 300, $\alpha=\beta=0.85$, $\alpha_f=\beta_f=0.9$ and $K=20$.}
\label{Fig-global}
\end{figure}

\begin{figure}[!t]
\centering
\includegraphics[width=1\linewidth]{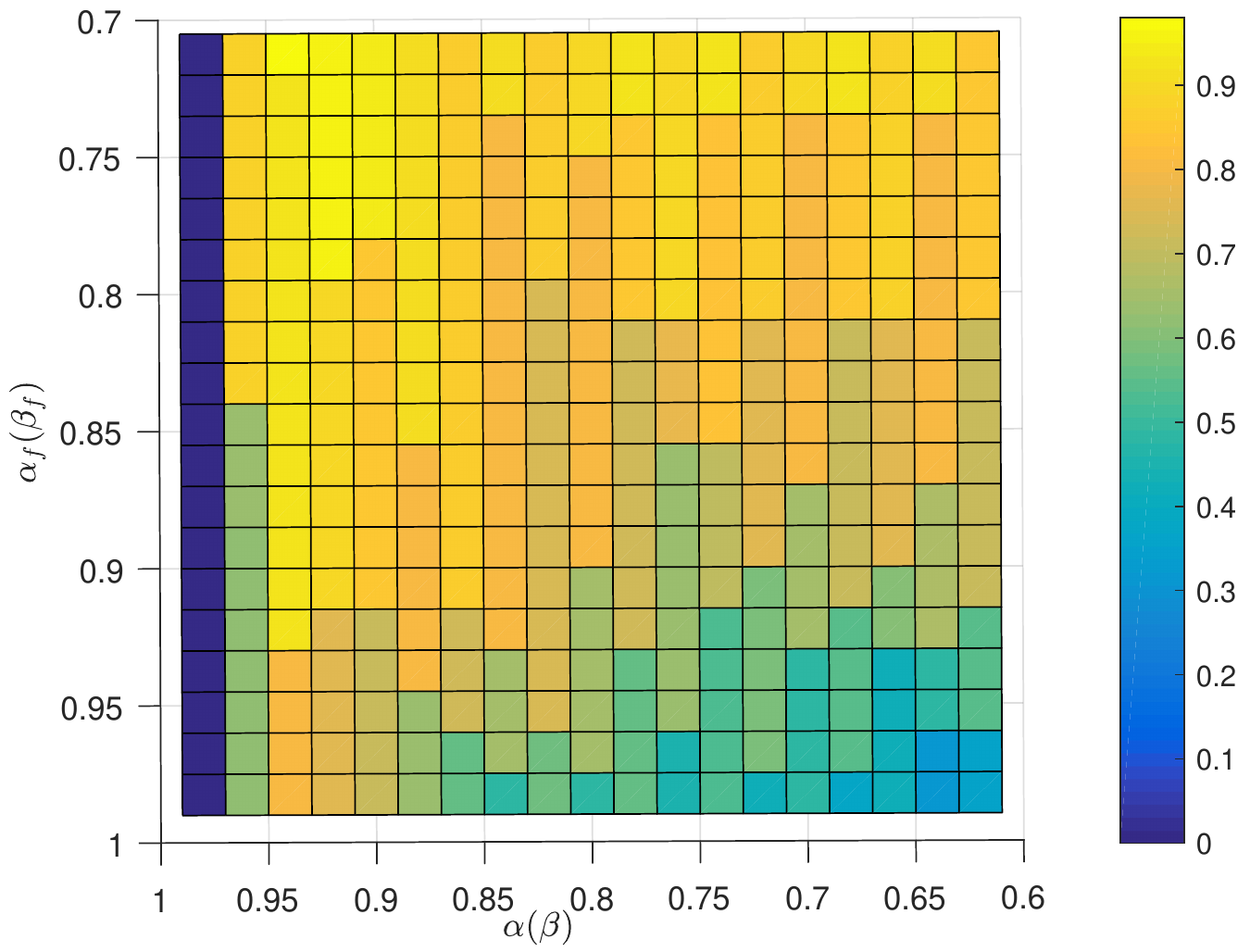}
\caption{ROC curves: $\Pr(\mathcal{H}_2|\mathcal{H}_2)$ versus the local constraints and global ones about $\Pr(\mathcal{H}_0|\mathcal{H}_0)$ and $\Pr(\mathcal{H}_1|\mathcal{H}_1)$, i.e., $(\alpha$, $\beta)$ and $(\alpha_f$, $\beta_f)$. Here, $\alpha=\beta$, $\alpha_f=\beta_f$, $N=300$, $K=20$, and the ratio of the received power from the IU to the noise power is -3.5dB.}
\label{Fig-roc-global}
\end{figure}
\begin{figure}[!t]
\centering
\includegraphics[width=1\linewidth]{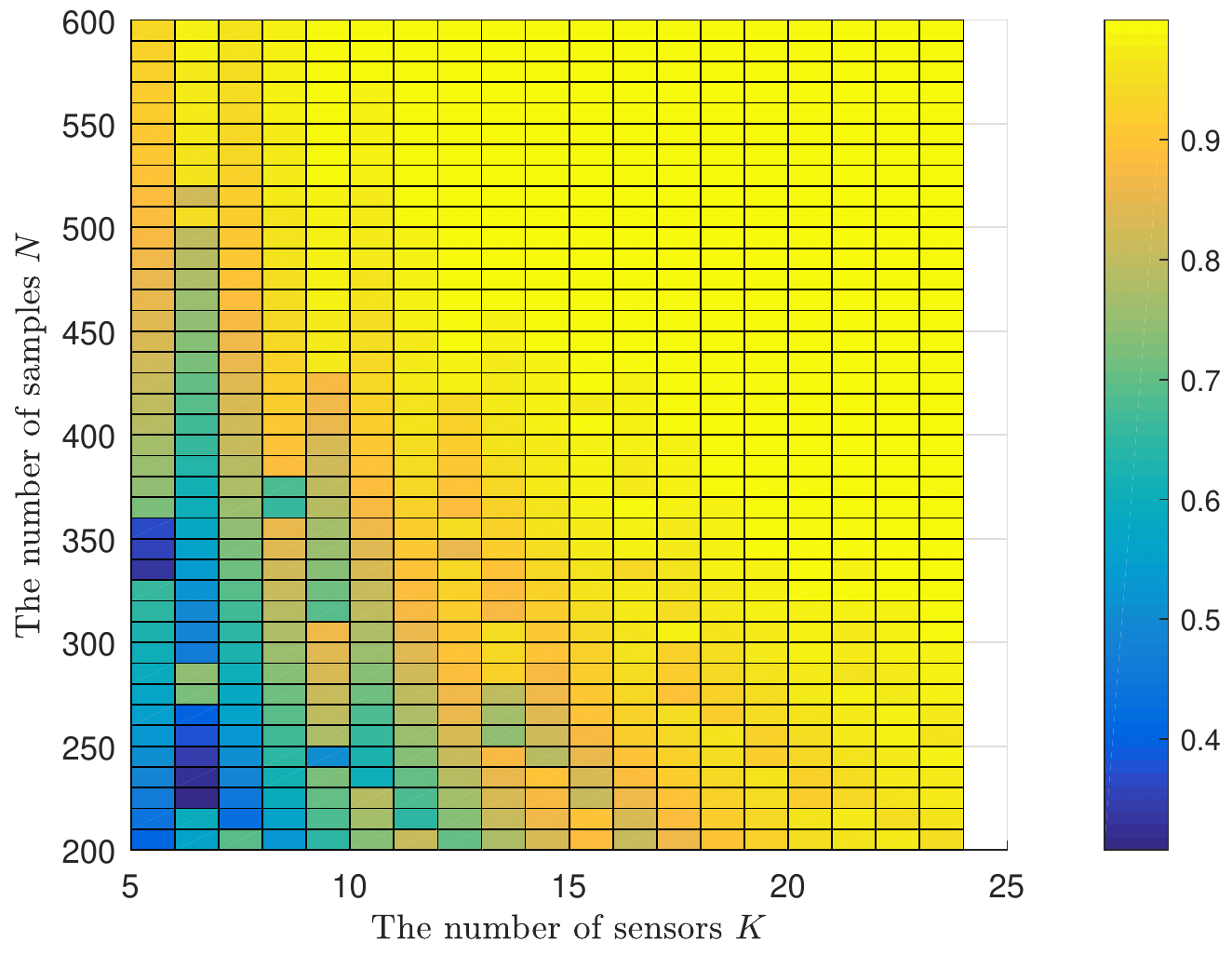}
\caption{The global detection probability $\Pr(\mathcal{H}_2|\mathcal{H}_2)$ versus the number of samples $N$ and the number of sensors $K$. Here $\alpha=\beta=0.85$, $\alpha_f=\beta_f=0.95$, and the ratio of the received power from the IU to the noise power is -3dB.}
\label{Fig-sample-nu-global}
\end{figure}

In Fig. \ref{Fig-global}, we find that cooperation between sensors can greatly improve the detection performance, which is true for detecting both the IU and the LU. Nevertheless, when $\sigma_2^2$ approaches $\sigma_0^2$ or $\sigma_1^2$, the detection probability decreases sharply, which is consistent to that when a single sensor makes the test. On the other side, due to the cooperation, the gaps between the schemes are narrowed and become almost zero, when $\sigma_2^2$ ranges from $1 \times 10^{-5}$ to $1.2 \times 10^{-5}$.

Fig. \ref{Fig-roc-global} shows that the global detection probability $\Pr(\mathcal{H}_2|\mathcal{H}_2)$ decreases with $\alpha_f$ and $\beta_f$, where the P-GLRT is exploited in each sensor. It is found that the cooperation between sensors dramatically decreases the uncertainty of detection. Simultaneously, for certain constraints $(\alpha_f,\beta_f)$ and a certain value of $\sigma_2^2$, there exist optimal local constraints $(\alpha,\beta)$ which are not too low or too high just as shown in Fig. \ref{Fig-roc-global}. However, the optimal solution of $(\alpha,\beta)$ is hard to obtain as it is related to the unknown variance $\sigma_2^2$.
Here it is needed to pointed out that when $\alpha$ and $\beta$ are set over 0.95, the local constraints cannot be satisfied any more, so we set the detection performance to zero.

Fig. \ref{Fig-sample-nu-global} shows that as the number of samples and the number of sensors increase, the global detection probability increases. Compared with Fig. \ref{Fig-sample-local}, much fewer samples are needed to achieve a favorable performance due to cooperation of multiple sensors. Further, increasing the number of samples is, in a certain degree, equivalent to increasing the number of sensors, where the former brings about a longer detection delay while the latter one leads to a higher hardware cost.
It is implied that we ought to make appropriate choices between the number of samples, i.e., the detection delay, and the number of sensors, i.e., the hardware cost, in order to optimize the resources and achieve the detection objective.

\subsection{Two-step Approximation and Its Optimality}
To further evaluate the global performance with multiple sensors, in particular, the effectiveness of the two-step approximation in the cooperative spectrum sensing (from Problem (49) to Problem (51), and from Problem (51) to Algorithm 1), three schemes are compared:
\begin{itemize}
  \item An ideal scheme named \emph{Oracle about (49)}, where the variance in $\mathcal{H}_2$, i.e., $\sigma_2^2$, is assumed to be known when the global decision is done. In this scheme, $\Pr (\mathbf{d}|{\mathcal{H}_2})$ can be calculated so that we can find the optimal solution to maximize $\Pr ({\mathcal{H}_2}|{\mathcal{H}_2})$  through exhaustive searching.
  \item The scheme of finding out the optimal solution about (51), named \emph{Optimal about (51)}. We can obtain the optimal solution for Problem (51) through exhaustive searching, where the number of elements in $S_2$ is maximized under the constraints.
  \item The heuristic scheme obtained by Algorithm 1, named \emph{Algorithm 1}.
\end{itemize}

Considering the closeness between the GLRT and Rao test in term of the detection performance, without loss of generality, in this subsection, GLRT is used in local decisions. In addition, it is pointed out that for exhaustive searching, the search space is huge and the computational complexity is high. Specifically, the number of possible combinations $[S_0,S_1,S_2]$ is $N_c=\sum\limits_{i = 1}^{L - 2} {\sum\limits_{j = 1}^{L - i - 1} {C_L^iC_{L - i}^j} }$, where $C_n^m = \frac{{n!}}{{m!(n - m)!}}$, which means that when the number of sensors $K$ is 5, for example, $N_c=1.0454\times10^{10}$. In contrast, the number of combinations searched, in Algorithm 1, is no more than $\frac{N}{2}$. From this perspective, the proposed heuristic algorithm is very necessary to reduce the computational complexity significantly. Considering the high computational complexity related to the number of sensors, the number of sensors are set to 5 in this subsection. Moreover, we find that there may exist multiple solutions about (51) which maximize $|S_2|$ while satisfying the constraints. Generally, the higher $\Pr(\mathcal{H}_0|\mathcal{H}_0)+\Pr(\mathcal{H}_1|\mathcal{H}_1)$ is, the lower $\Pr(\mathcal{H}_2|\mathcal{H}_2)$ is, which is identical to part of the motivation of obtaining (50) from (49). Hence, when this case happens, the solution which maximizes $\Pr(\mathcal{H}_0|\mathcal{H}_0)+\Pr(\mathcal{H}_1|\mathcal{H}_1)$ is chosen.

\begin{figure}[!t]
\centering
\includegraphics[width=1\linewidth]{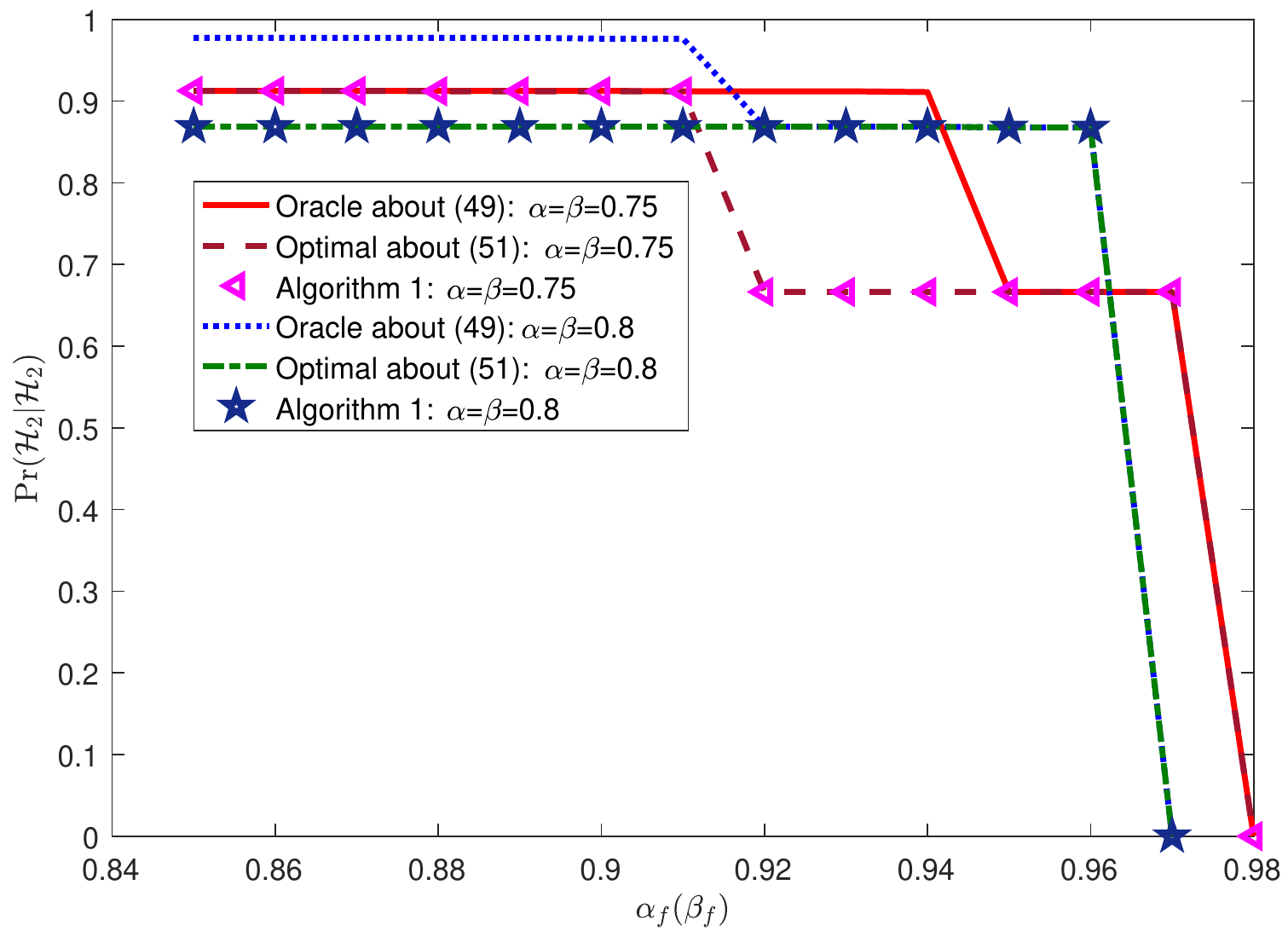}
\caption{$\Pr ({\mathcal{H}_2}|{\mathcal{H}_2})$ versus the constraints $\alpha_f$ and $\beta_f$ under different schemes. Here, $\sigma_1^2=1.2\times 10^{-5}$ Watt, $\sigma_2^2=1.4\times10^{-5}$ Watt, and $K=5$.}
\label{Fig-gap}
\end{figure}

Fig. \ref{Fig-gap} depicts $\Pr ({\mathcal{H}_2}|{\mathcal{H}_2})$ versus the constraints $\alpha_f$ and $\beta_f$ and shows the gaps between different schemes under different local constraints. As expected, the detection performance decreases with the constraints. As the process of global decision is an integer linear programming, the curve is non-smooth. When $\alpha_f$ and $\beta_f$ reaches some level, they cannot be satisfied any more and $\Pr ({\mathcal{H}_2}|{\mathcal{H}_2})$ is set to zero. We find that \emph{Algorithm 1} works closely with \emph{Optimal about (51)} and \emph{Oracle about (49)} is better than the other two at certain intervals and keeps similar performance at other intervals.
The local constraints work through controlling the probability distributions of $\mathbf{d}$.
The higher the local constraints are, the more intensive the probability distributions under ${\mathcal{H}_0}$ and ${\mathcal{H}_1}$ are and the more dispersive the probability distribution under ${\mathcal{H}_2}$ is. Hence, when the constraints changes from 0.75 to 0.8, the schemes encounter relatively smooth declines with the increasing global constraints.

\section{Conclusion}
In this paper, we investigated the problem on the detection of spectrum misuse behaviors brought by illegitimate access or rogue power emission. To detect whether the channel is occupied and recognize whether the illegitimate user exists, we exploited the multi-hypothesis test to model the spectrum sensing problem under the spectrum misuse behaviors, where the states with the existence of the IU were formulated as composite hypotheses due to the unknown characteristic of the IU. We built a test criterion called the generalized multi-hypothesis Neyman-Pearson (GMNP) criterion and derived two test rules based on the GLRT and the Rao test, respectively. In particular, we analyzed the problem of overlapping between decision regions raised by the multi-hypothesis characteristic and rigid constraints. To evaluate the test rule, the asymptotic performance was derived and an upper bound of the detection performance was also given through introducing the prior knowledge about the range of the unknown variance. Furthermore, for the multi-hypothesis test problem, a cooperative spectrum sensing scheme was developed based on the global GMNP criterion. The simulation results verified the detection performances in terms of the unknown variance, the number of the samples, the constraints, and the number of sensors.

\begin{appendices}
\section{Proof of Lemma 1}
Let $\mathcal{A}$ denote the space of the observation vectors, that is, $\mathcal{A}=\{\textbf{y}:{y_i} \in \textbf{R},i=0,1,...,N\}$.
Then, ${\mathcal{R}_0}$, ${\mathcal{R}_1}$, and ${\mathcal{R}_2}$ are complete and mutually exclusive sets, that is, ${\mathcal{R}_0} \cup {\mathcal{R}_1} \cup {\mathcal{R}_2} = \mathcal{A}$, ${\mathcal{R}_i} \cap {\mathcal{R}_j} = \emptyset ,i \ne j$.
As $\Pr ({\mathcal{H}_i}|{\mathcal{H}_i})=\int_{{\mathcal{R}_i}} {p(y;{\mathcal{H}_i})} dy$, to decrease $\Pr ({\mathcal{H}_i}|{\mathcal{H}_i})$ is equivalent to shrink $\mathcal{R}_i$. Therefore, if $\Pr ({\mathcal{H}_0}|{\mathcal{H}_0})$ and $\Pr ({\mathcal{H}_1}|{\mathcal{H}_1})$ decrease, both $\mathcal{R}_0$ and $\mathcal{R}_1$ shrink. As $\mathcal{R}_2=\mathcal{A}-\mathcal{R}_0-\mathcal{R}_1$, $\mathcal{R}_2$ is enlarged and $\Pr ({\mathcal{H}_2}|{\mathcal{H}_2})$ increases.

Hence, when the maximum of $\Pr ({\mathcal{H}_2}|{\mathcal{H}_2})$ is achieved, $\mathcal{R}_0$ and $\mathcal{R}_0$ are compressed so that the conditions are justly satisfied, i.e., $\Pr ({\mathcal{H}_0}|{\mathcal{H}_0}) = \alpha$, $\Pr ({\mathcal{H}_1}|{\mathcal{H}_1}) = \beta$. Hence, we obtain the equivalent optimization problem where the inequalities are substituted with the equalities.
\section{Proof of Theorem 1}
Firstly, we consider the critical point $(\alpha_c, \beta_c)$ which makes the overlapping just happens, i.e., $\eta_0=\eta_1=\eta_1^{*}$. First, based on (\ref{Eq_eta0}), $\eta_1^{*}$ is formulated
\[\eta_1^{*} = {\Gamma ^{ - 1}}\left(\frac{N}{2},(1-\alpha_c) \Gamma (\frac{N}{2})\right)\sigma _0^2.\]

Substitute $\eta_1^{*}$ for $\eta_1$ in (\ref{Eq_threshold1}), and we have
\[{(\frac{{\eta_1^{*}}}{{\eta _2}})^{\frac{N}{2}}}\exp (\frac{{\eta _2 - \eta_1^{*}}}{{2\sigma _1^2}}) = 1.\]
So, we can obtain $\eta _2^{*}$ that is the solution to the equation above, $\eta _2\neq \eta_1$. Then, $\Pr(\mathcal{H}_1|\mathcal{H}_1)$ is formulated as follows
\begin{equation}
{\beta _c} = \frac{{\Gamma (\frac{N}{2},\frac{{{\eta_1^{*}}}}{{\sigma _1^2}})}}{{\Gamma (\frac{N}{2})}} - \frac{{\Gamma (\frac{N}{2},\frac{{{\eta _2^{*}}}}{{\sigma _1^2}})}}{{\Gamma (\frac{N}{2})}}.
\label{Eq_theo2}
\end{equation}

Further, it is found that when $\beta$ increases, $\lambda_1$ increases, $\eta_1$ decreases, and the area of the overlapping region increases. On the other side, when $\alpha$ increases, $\eta_0$ decreases and the area of the overlapping region increases. This means that once $\beta$ is over the value in (\ref{Eq_theo2}), the overlapping happen. In conclusion, the overlapping condition is obtained just as formulated in (\ref{Eq_overlapping}).

\section{Proof of Theorem 2}
Similar to the proof of Theorem 1, we find the critical point $(\alpha_c^r, \beta_c^r)$ which makes $\eta_0^R=\eta_1^R=\eta_1^r$.
First, based on (\ref{Eq_eta0}), we have
\[\eta_0^r = {\Gamma ^{ - 1}}\left(\frac{N}{2},(1-\alpha_c^r) \Gamma (\frac{N}{2})\right)\sigma _0^2.\]
Based on (\ref{Eq_rao}), we have $\eta_2^{r} =2N\sigma_1^2-\eta_1^r$. Further, when $\beta$ increases, $\lambda_1^R$ increases, $\eta_1^R$ decreases, and the area of the overlapping region increases. On the other side, when $\alpha$ increases, $\eta_0^R$ decreases and the area of the overlapping region increases. Hence, we obtain the overlapping condition in (\ref{Eq_overrao}).
\section{Proof of Theorem 3}
In (\ref{Eq_optima}), we obtain that to maximize $\Pr(\mathcal{H}_2|\mathcal{H}_2)$, in the decision regions ${\mathcal{R}_0}$ and ${\mathcal{R}_1}$, we have
\begin{equation}
\left\{ {\begin{array}{*{20}{c}}
{\frac{{p({\mathbf{y}};{{\mathcal {H}}_2})}}{{p({\mathbf{y}};{{\mathcal {H}}_0})}} < {\lambda _0},}&{{{\mathcal {R}}_0},}\\
{\frac{{p({\mathbf{y}};{{\mathcal {H}}_2})}}{{p({\mathbf{y}};{{\mathcal {H}}_1})}} < {\lambda _0},}&{{{\mathcal {R}}_1}.}
\end{array}} \right.
\label{Eq_R3}
\end{equation}

Then, based on the range of $\sigma_2^2$, two cases are considered respectively.

Case 1:  $\sigma_0^2<\sigma_2^2<\sigma_1^2$. Based on (\ref{Eq_R3}), we have
\[\left\{ {\begin{array}{*{20}{c}}
  {Y <  \frac{2{\sigma _0^2\sigma _2^2}}{{\sigma _2^2 - \sigma _0^2}}(\ln{\lambda _0}-\frac{N}{2}\ln (\frac{{\sigma _2^2}}{{\sigma _0^2}})) = {\eta _0},}&{{{\mathcal {R}}_0},}\\
  {Y > \frac{2{\sigma _1^2\sigma _2^2}}{{\sigma _2^2 - \sigma _1^2}}(\ln{\lambda _1}-\frac{N}{2}\ln (\frac{{\sigma _2^2}}{{\sigma _1^2}})) = {\eta _1},}&{{{\mathcal {R}}_1}.}
\end{array}} \right. \]
Due to lack of the a prior knowledge of $\sigma_2^2$, it seems to be hard to build the detector. However, under the hypothesis $\mathcal{H}_i$, the statistic $\frac{Y}{\sigma_i^2} \sim \chi _N^2$. Therefore,

\[\left\{ \begin{gathered}
  \Pr(\mathcal{H}_0|\mathcal{H}_0) = \int_0^{{\eta _0}} {p(Y;{\mathcal{H}_0})dY}=1 - \frac{{\Gamma (\frac{N}{2},\frac{{{\eta _0}}}{{\sigma _0^2}})}}{{\Gamma (\frac{N}{2})}}   = \alpha,  \hfill \\
  \Pr(\mathcal{H}_1|\mathcal{H}_1) = \int_{{\eta _1}}^{ + \infty } {p(Y;{\mathcal{H}_1})dY}=\frac{{\Gamma (\frac{N}{2},\frac{{{\eta _1}}}{{\sigma _1^2}})}}{{\Gamma (\frac{N}{2})}}   = \beta.  \hfill \\
\end{gathered}  \right.\]
Hence, without the a prior knowledge of $\sigma_2^2$, we obtain the two thresholds, $\eta _0 = {\Gamma ^{ - 1}}(\frac{N}{2},(1-\alpha) \Gamma (\frac{N}{2}))\sigma _0^2$, $\eta _1 = {\Gamma ^{ - 1}}(\frac{N}{2},\beta \Gamma (\frac{N}{2}))\sigma _1^2$.

Case 2: $\sigma_2^2 > \sigma_1^2$
Similarly, we have
\[\left\{ {\begin{array}{*{20}{c}}
  {Y <  \frac{2{\sigma _0^2\sigma _2^2}}{{\sigma _2^2 - \sigma _0^2}}(\ln{\lambda _0}-\frac{N}{2}\ln (\frac{{\sigma _2^2}}{{\sigma _0^2}})) = {\eta _0},}&{{{\mathcal {R}}_0},}\\
  {Y < \frac{2{\sigma _1^2\sigma _2^2}}{{\sigma _2^2 - \sigma _1^2}}(\ln{\lambda _1}-\frac{N}{2}\ln (\frac{{\sigma _2^2}}{{\sigma _1^2}})) = {\eta _1},}&{{{\mathcal {R}}_1}.}
\end{array}} \right. \]

As $\sigma_0^2 < \sigma_1^2$, the two constraints are reformulated as
\[\left\{ \begin{gathered}
  \Pr ({\mathcal{H}_0}|{\mathcal{H}_0})= 1-\frac{{\Gamma (\frac{N}{2},\frac{{{\eta _0}}}{{\sigma _0^2}})}}{{\Gamma (\frac{N}{2})}}  = \alpha,  \hfill \\
  \Pr ({\mathcal{H}_1}|{\mathcal{H}_1}) = \frac{{\Gamma (\frac{N}{2},\frac{{{\eta _0}}}{{\sigma _1^2}})}}{{\Gamma (\frac{N}{2})}}- \frac{{\Gamma (\frac{N}{2},\frac{{{\eta _1}}}{{\sigma _1^2}})} }{{\Gamma (\frac{N}{2})}} = \beta.  \hfill \\
\end{gathered}  \right.\]
Hence, we obtain the two thresholds, $\eta _0 = {\Gamma ^{ - 1}}(\frac{N}{2},(1-\alpha) \Gamma (\frac{N}{2}))\sigma _0^2$, $\eta_1={\Gamma ^{ - 1}}(\frac{N}{2},\Gamma (\frac{N}{2},\frac{{{\eta _0}}}{{\sigma _1^2}}) - \beta \Gamma (\frac{N}{2}))\sigma_1^2$.
\end{appendices}

\begin{IEEEbiography}[{\includegraphics[width=1in,height=1.25in,clip,keepaspectratio]{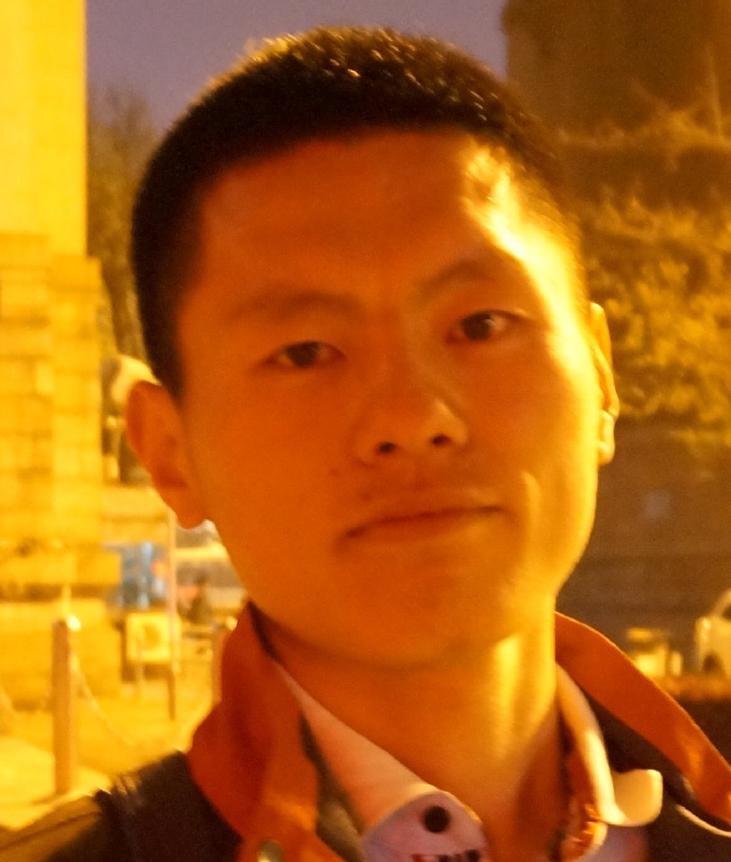}}]{Linyuan Zhang}
received his B.S. degree (with honors) in electronic engineering from Inner Mongolia University, Hohhot, China, in 2012. He is currently pursuing his Ph.D. degree in communications and information system in College of Communications Engineering, Army Engineering University of PLA. His research interests are wireless security and statistical learning.
\end{IEEEbiography}
\begin{IEEEbiography}[{\includegraphics[width=1in, height=1.25in,clip,keepaspectratio]{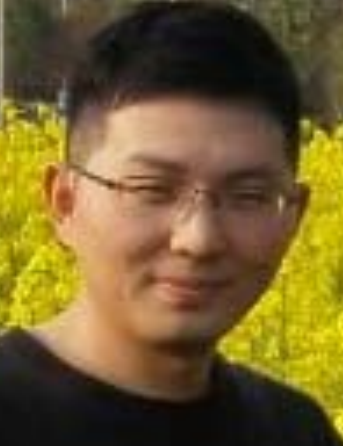}}]{Guoru Ding} (S'10-M'14-SM'16) received his B.S. degree (Hons.) in electrical engineering from Xidian University, Xi'an, China, in 2008 and his Ph.D. degree (Hons.) in communications and information systems in College of Communications Engineering, Nanjing, China, in 2014. Since 2014, he has been an assistant professor in College of Communications Engineering and a research fellow in National High Frequency Communications Research Center of China. Since April 2015, he has been a Postdoctoral Research Associate at the National Mobile Communications Research Laboratory, Southeast University, Nanjing, China. His research interests include cognitive radio networks, massive MIMO, machine learning, and big data analytics over wireless networks.

He has served as a Guest Editor of the \textsc{IEEE Journal on Selected Areas in Communications} (Special issue on spectrum sharing and aggregation in future wireless networks). He is now an Associate Editor of \textsc{the Journal of Communications and Information Networks}, \textsc{the KSII Transactions on Internet and Information Systems} and \textsc{the AEU-International Journal of Electronics and Communications}. He has acted as Technical Program Committees (TPC) members for a number of international conferences, including the IEEE Global Communications Conference (GLOBECOM), IEEE International Conference on Communications (ICC), and IEEE Vehicular Technology Conference (VTC). He is a Voting Member of the IEEE 1900.6 Standard Association Working Group. He was a recipient of the Best Paper Awards from EAI MLICOM 2016, IEEE VTC 2014-Fall, and IEEE WCSP 2009. He was awarded the Alexander von Humboldt Fellowship in 2017 and the Excellent Doctoral Thesis Award of China Institute of Communications in 2016.
\end{IEEEbiography}

\begin{IEEEbiography}[{\includegraphics[width=1in,height=1.25in,clip,keepaspectratio]{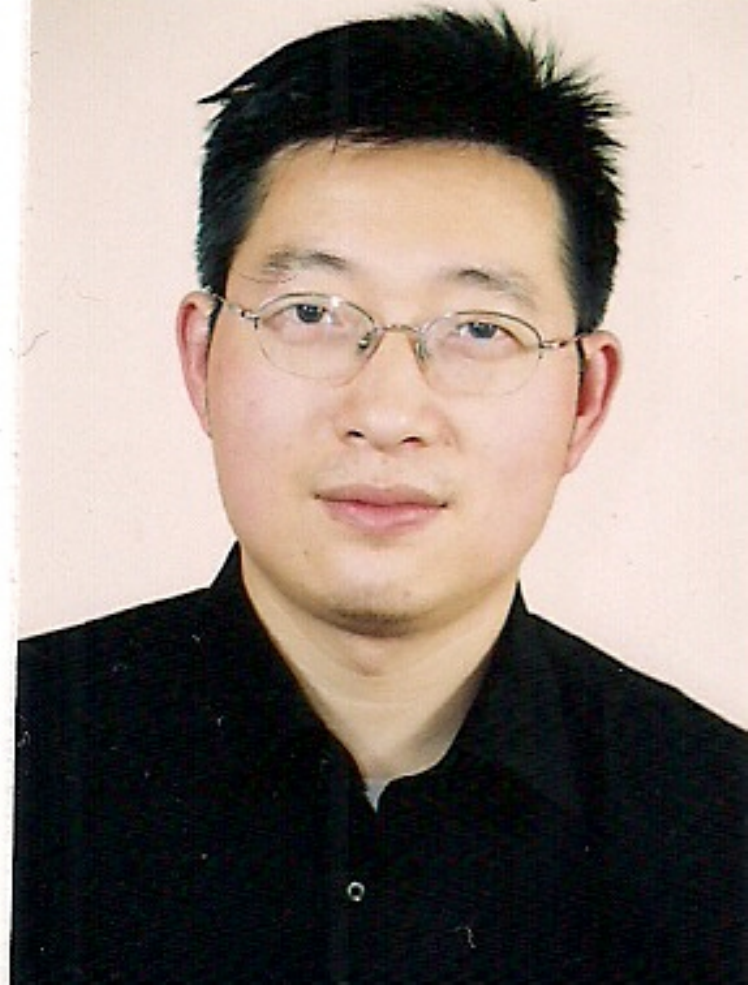}}]{Qihui Wu} (SM'13) received his B.S. degree in communications engineering, M.S. degree and Ph.D. degree in communications and information systems from Institute of Communications Engineering, Nanjing, China, in 1994, 1997 and 2000, respectively. From 2003 to 2005, he was a Postdoctoral Research Associate at Southeast University, Nanjing, China. From 2005 to 2007, he was an Associate Professor with the College of Communications Engineering, PLA University of Science and Technology, Nanjing, China, where he served as a Full Professor from 2008 to 2016. Since May 2016, he has been a full professor with the College of Electronic and Information Engineering, Nanjing University of Aeronautics and Astronautics, Nanjing, China. From March 2011 to September 2011, he was an Advanced Visiting Scholar in Stevens Institute of Technology, Hoboken, USA. Dr. Wu's current research interests span the areas of wireless communications and statistical signal processing, with emphasis on system design of software defined radio, cognitive radio, and smart radio.
\end{IEEEbiography}
\begin{IEEEbiography}[{\includegraphics[width=1in,height=1.25in,clip,keepaspectratio]{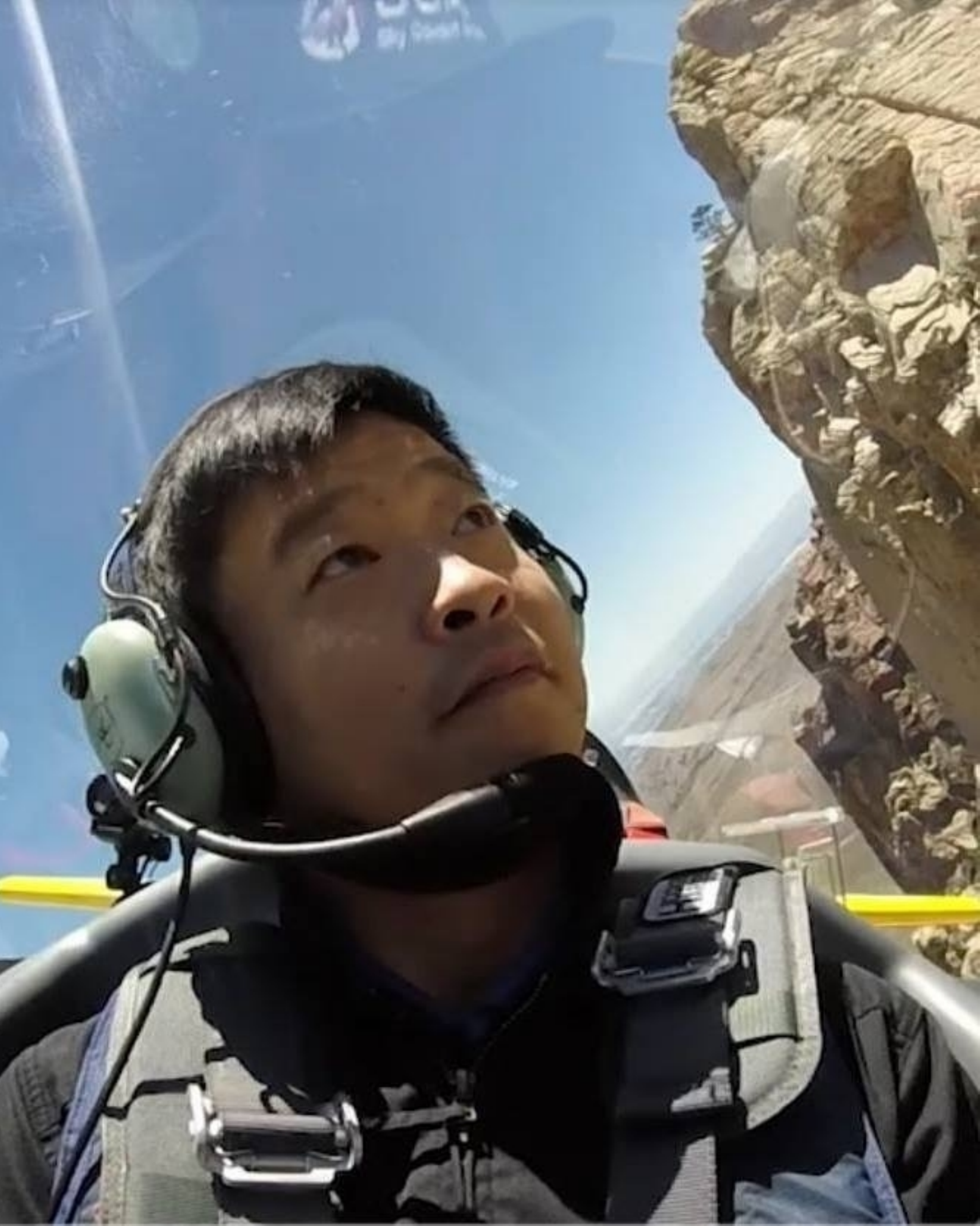}}]{Zhu Han}
(S'01-M'04-SM'09-F'14) received the B.S. degree in electronic engineering from Tsinghua University, in 1997, and the M.S. and Ph.D. degrees in electrical and computer engineering from the University of Maryland, College Park, in 1999 and 2003, respectively.

From 2000 to 2002, he was an R\&D Engineer of JDSU, Germantown, Maryland. From 2003 to 2006, he was a Research Associate at the University of Maryland. From 2006 to 2008, he was an assistant professor at Boise State University, Idaho. Currently, he is a Professor in the Electrical and Computer Engineering Department as well as in the Computer Science Department at the University of Houston, Texas. His research interests include wireless resource allocation and management, wireless communications and networking, game theory, big data analysis, security, and smart grid.  Dr. Han received an NSF Career Award in 2010, the Fred W. Ellersick Prize of the IEEE Communication Society in 2011, the EURASIP Best Paper Award for the Journal on Advances in Signal Processing in 2015, IEEE Leonard G. Abraham Prize in the field of Communications Systems (best paper award in IEEE JSAC) in 2016, and several best paper awards in IEEE conferences. Currently, Dr. Han is an IEEE Communications Society Distinguished Lecturer.
\end{IEEEbiography}


\begin{thebibliography}{99}
\bibitem{Mitola}
J. Mitola and G. Q. Maguire, ``Cognitive radio: Making software radios more personal,'' \emph{IEEE Personal Communications}, vol. 6, no. 4, pp. 13-18, Aug. 1999.
\bibitem{Dynamic_spectrum}
E. Hossain, D. Niyato, and Z. Han, \emph{Dynamic Spectrum Access and Management in Cognitive Radio Networks}, Cambridge University Press, 2009.
\bibitem{Haykin}
S. Haykin, ``Cognitive radio: Brain-empowered wireless communications,'' \emph{IEEE Journal on Selected Areas in Communications}, vol. 23, no. 2, pp. 201-220, Feb. 2005.


\bibitem{CST}
L. Zhang, G. Ding, Q. Wu, Y. Zou, Z. Han, and J. Wang, ``Byzantine attack and defense in cognitive radio networks: A survey,'' \emph{IEEE Communications Surveys and Tutorials}, vol. 17, no. 3, pp. 1342-1363, Third quarter 2015.

\bibitem{Ding_Tcom}
G. Ding, J. Wang, Q. Wu, L. Zhang, Y. Zou, Y. D. Yao, and Y. Chen, ``Robust spectrum sensing with crowd sensors,'' \emph{IEEE Transaction on Communications}, vol. 62, no. 9, pp. 3129-3143, Sep. 2014.




\bibitem{Chen_2008}
R. Chen, J. M. Park, and J. H. Reed, ``Defense against primary user emulation attacks in cognitive radio networks,'' \emph{IEEE Journal on Selected Areas in Communications}, vol. 26, no. 1, pp. 25-37, Mar. 2008.

\bibitem{GLobecom_2012}
A. Garnaev, W. Trappe, and C. T. Kung, ``Dependence of optimal monitoring strategy on the application to be protected,'' in \emph{2012 IEEE Global Communications Conference (GLOBECOM)}, Anaheim, CA, 3-7 Dec. 2012, pp. 1054-1059.

\bibitem{Dogfight_1}
H. Li and Z. Han, ``Dogfight in spectrum: Combating primary user emulation attacks in cognitive radio systems-Part I: Known channel statistics,'' \emph{IEEE Transactions on Wireless Communications}, vol. 9, no. 11, pp. 3566-3577, Sep. 2010.

\bibitem{Dogfight_2}
H. Li and Z. Han, ``Dogfight in spectrum: Combating primary user emulation attacks in cognitive radio systems-Part II: Unknown channel statistics, '' \emph{IEEE Transactions on Wireless Communications}, vol. 10, no. 1, pp. 274-283, Dec. 2011.
\bibitem{Belief_PUE}
Z. Yuan, D. Niyato, H. Li, J. B. Song, and Z. Han, ``Defeating primary user emulation attacks using belief propagation in cognitive radio networks,'' \emph{IEEE Journal on Selected Areas in Communications}, vol. 30, no. 10, pp. 1850-1860, Nov. 2012.
\bibitem{nonpa_Bayesian}
N. T. Nguyen, R. Zheng, and Z. Han, ``On identifying primary user emulation attacks in cognitive radio systems using nonparametric Bayesian classification,'' \emph{IEEE Transactions on Signal Processing}, vol. 60, no. 3, pp. 1432-1445, Mar. 2012.

\bibitem{ALDO}
S. Liu, Y. Chen, W. Trappe, and L. J. Greenstein, ``ALDO: An anomaly detection framework for dynamic spectrum access networks,'' in \emph{2009 IEEE Conference on Computer Communications (INFOCOM)}, Rio de Janerio, Brazil, 19-25 Apr. 2009, pp. 675-683. 
\bibitem{PUEA_model}
V. Kumar, J. M. J. Park, and K. Bian,, ``PHY-layer authentication using duobinary signaling for spectrum enforcement,'' \emph{IEEE Transactions on Information Forensics and Security}, vol. 11, no. 5, pp. 1027-1038, May 2016.
\bibitem{Clock_Skews}
S. Jana and S. K. Kasera, ``On fast and accurate detection of unauthorized wireless access points using clock skews,'' \emph{IEEE Transactions on Mobile Computing}, vol. 9, no. 3, pp. 449-462, Aug. 2010.
\bibitem{Permits_mobihoc}
L. Yang, Z. Zhang, B. Y. Zhao, C. Kruegel, and H. Zheng, ``Enforcing dynamic spectrum access with spectrum permits,'' in \emph{the Thirteenth ACM  International Symposium on Mobile Ad Hoc Networking and Computing (MobiHoc)}, Hilton Head Island, SC, 11–14 Jun. 2012, pp. 195-204.

\bibitem{Lin_access}
L. Zhang, G. Ding, Q. Wu, and F. Song, ``Defending against Byzantine attack in cooperative spectrum sensing: Defense reference and performance analysis,'' \emph{IEEE Access}, vol. 4, pp. 4011-4024, Apr. 2016.

\bibitem{Nie_access}
G. Nie, G. Ding, L. Zhang, and Q. Wu, ``Byzantine defense in collaborative spectrum sensing via Bayesian learning," \emph{IEEE Access}, vol. 5, pp. 20089-20098, Sep. 2017.

\bibitem{Attack_preve}
L. Duan, A. W. Min, J. Huang, and K. G. Shin, ``Attack prevention for collaborative spectrum sensing in cognitive radio networks,'' \emph{IEEE Journal on Selected Areas in Communications}, vol. 30, no. 9, pp. 1658-1665, Oct. 2012.

\bibitem{Thwart}
W. Wang, L. Chen, K. G. Shin, and L. Duan, ``Thwarting intelligent malicious behaviors in cooperative spectrum sensing,'' \emph{IEEE Transactions on Mobile Computing}, vol. 14, no. 11, pp. 2392-2405, Nov. 2015.

\bibitem{See_something}
A. Dutta and M. Chiang, ```See something, say something' crowdsourced enforcement of spectrum policies,'' \emph{IEEE Transactions on Wireless Communications}, vol. 15, no. 1, pp. 67-80, Jan. 2016.
\bibitem{SpecWathch}
M. Li, D. Yang, J. Lin, M. Li, and J. Tang, ``SpecWatch: Adversarial spectrum usage monitoring in CRNs with unknown statistics,'' in \emph{the 35th Annual IEEE International Conference on Computer Communications (INFOCOM)}, San Francisco, CA, 10-14 Apr. 2016, pp. 1-9.

\bibitem{rogue_mass2012}
K. Tan, K. Zeng, D. Wu, and P. Mohapatra, ``Detecting spectrum misuse in wireless networks,'' in \emph{2012 IEEE 9th International Conference on Mobile Ad-Hoc and Sensor Systems (MASS 2012)}, Las Vegas, NV, 8-11 Oct. 2012, pp. 245-253.

\bibitem{Yao_css}
C. Chen, H. Cheng, and Y D. Yao, ``Cooperative spectrum sensing in cognitive radio networks in the presence of the primary user emulation attack,'' \emph{IEEE Transactions on Wireless Communications}, vol. 10, no. 7, pp. 2135-2141, Apr. 2011.

\bibitem{Three_fingerpints}
N. Gao, X. Jing, H. Huang, and J. Mu, ``Robust collaborative spectrum sensing using PHY-Layer fingerprints in mobile cognitive radio networks,'' \emph{IEEE Communications Letters}, vol. 21, no. 5, pp. 1063-1066, May 2017.

\bibitem{Ad_hoc}
S. Liu, L. J. Greenstein, W. Trappe, and Y. Chen, ``Detecting anomalous spectrum usage in dynamic spectrum access networks,'' \emph{Ad Hoc
Networks}, vol. 10, no. 5, pp. 831–844, Jul. 2012.



\bibitem{Gaofeifei}
F. Gao, J. Li, T. Jiang, and W. Chen, ``Sensing and recognition when primary user has multiple transmit power levels,'' \emph{IEEE Transactions on Signal Processing}, vol. 63, no. 10, pp. 2704-2717, Mar. 2015.


\bibitem{Sequ_multi}
Z. Li, S. Cheng, F. Gao, and Y. C. Liang, ``Sequential detection for cognitive radio with multiple primary transmit power levels,'' \emph{IEEE Transactions on Communications}, vol. 65, no. 7, pp. 2769-2780, Jul. 2017.

\bibitem{Genera_Rao}
D. Ciuonzo, P. Salvo Rossi, and P. Willett, ``Generalized rao test for decentralized detection of an uncooperative target,'' \emph{IEEE Signal Processing Letters}, vol. 24, no. 5, pp. 678-682, Mar. 2017.
\bibitem{GML}
S. M. Kay, \emph{Fundamentals of Statistical Signal Processing: Detection Theory}, Upper Saddle River, NJ: Prentice-Hall, 1998.

\bibitem{spectrum_sensing}
T. Yucek and H. Arslan, ``A survey of spectrum sensing algorithms for cognitive radio applications,'' \emph{IEEE Communications Surveys and Tutorials}, vol. 11, no. 1, pp. 116-130, First Quarter 2009.

\bibitem{sens_throu_trad}
Y. C. Liang, Y. Zeng, E. C. Y. Peh, and A. T. Hoang, ``Sensing-throughput tradeoff for cognitive radio networks,'' \emph{IEEE Transactions on Wireless Communications}, vol. 7, no. 4, pp. 1326-1337, Apr. 2008.

\bibitem{NP_tra}
F. L. Lehmann, \emph{Testing Statistical Hypotheses}, New York: Wiley, 1959.
\bibitem{CSS_zhangwei}
W. Zhang, R. K. Mallik, and K. B. Letaief, ``Optimization of cooperative spectrum sensing with energy detection in cognitive radio networks,'' \emph{IEEE Transactions on Wireless Communications}, vol. 8, no. 12, pp. 5761-5766, Dec. 2009.
\bibitem{Optimal_CSS}
Z. Quan, S. Cui, and A. Sayed, ``Optimal linear cooperation for spectrum sensing in cognitive radio networks,'' \emph{IEEE Journal on Selected Areas in Communications}, vol. 2, no. 1, pp. 28-40, Feb. 2008.
\bibitem{ieee802_22}
\emph{IEEE 802.22 Working Group on Wireless Regional Area Networks}, http://standards.ieee.org/about/get/802/802.22.html, 2015.





\end{thebibliography}
\end{document}